\documentclass[12pt]{iopart}

\expandafter\let\csname equation*\endcsname\relax
\expandafter\let\csname endequation*\endcsname\relax
\usepackage{mathrsfs}
\usepackage{amsmath,amsfonts,amssymb}

\usepackage{graphicx,psfrag}
\usepackage{multirow,enumerate}
\usepackage{comment,hyperref}
\usepackage{color}
\usepackage{acronym}
\usepackage{xspace}
\usepackage[normalem]{ulem}

\newacro{BH}{black hole}
\newacro{NS}{neutron star}
\newacro{PN}{Post-Newtonian}
\newacro{BBH}{binary black hole}
\newacro{BNS}{binary neutron star}
\newacro{EOB}{effective-one-body}
\newacro{NR}{numerical relativity}
\newacro{GW}{gravitational wave}
\newacro{EOS}{equation-of-state}

\def\BAM{{\texttt{BAM}}\xspace}

\def\GMc2{{\rm G M_{\odot} c^{-2}}}

\def\R{\mathcal{R}}
\def\I{\mathcal{I}}

\usepackage{color}

\definecolor{cyan}{rgb}{0,0.9,0.9}
\definecolor{orange}{rgb}{0.9,0.5,0}
\definecolor{magenta}{rgb}{1,0,1}
\definecolor{purple}{rgb}{0.8,0.4,0.8}
\definecolor{gray}{rgb}{0.5,0.5,0.5}
\definecolor{mygreen}{rgb}{0.1,0.8,0.1}
\definecolor{darkblue}{rgb}{0.0,0.0,0.6}

\begin{document}

\title[Full 3D NR Simulations of NS--BS Collisions with BAM]
{Full 3D Numerical Relativity Simulations of Neutron Star -- Boson Star Collisions with BAM}

\author{Tim Dietrich}
\address{Nikhef, Science Park, 1098 XG Amsterdam, The Netherlands}

\author{Serguei Ossokine}
\address{Max Planck Institute for Gravitational Physics (Albert Einstein Institute),
                    Am M\"uhlenberg 1, Potsdam 14476, Germany}

\author{Katy Clough}
\address{Institut f\"ur Astrophysik, Georg-August Universit\"at,
                    Friedrich-Hund-Platz 1, D-37077 G\"ottingen, Germany}

\date{\today}

\begin{abstract}
With the first direct detections of gravitational waves (GWs) from the coalescence of
compact binaries observed by the advanced LIGO and VIRGO interferometers,
the era of GW astronomy has begun. Whilst there is strong evidence that
the observed GWs are connected to the merger of two black holes (BH)
or two neutron stars (NS), future detections may present a less consistent picture.
Indeed, the possibility that the observed GW signal was created by a merger of
exotic compact objects (ECOs) such as boson stars (BS) or axion stars (AS) has
not yet been fully excluded.
For a detailed understanding of the late stages of the coalescence
full 3D numerical relativity simulations are essential.
In this paper, we extend the infrastructure of the numerical relativity
code \BAM, to permit the simultaneous simulation of baryonic
matter with bosonic scalar fields, thus enabling the study of BS-BS, 
BS-NS, and BS-BH mergers.
We present a large number of single star evolutions to test the newly implemented routines, and
to quantify the numerical challenges of such simulations, which we find to partially differ from the default NS case.
We also compare head-on BS-BS simulations with independent
numerical relativity codes, namely the SpEC and the GRChombo codes, and
find good general agreement.
Finally, we present what are, to the best of our knowledge, the first full NR simulations
of BS-NS mergers, a first step towards identifying the hallmarks
of BS-NS interactions in the strong gravity regime, as well as possible GW and electromagnetic
observables.
\end{abstract}

\maketitle

\section{Introduction}

The Advanced LIGO interferometers inaugurated the era of
gravitational wave (GW) astronomy with the first direct detection of
GWs in 2015~\cite{Abbott:2016blz}.
Since then numerous detections have been
made~\cite{Abbott:2016nmj,Abbott:2017vtc,Abbott:2017oio,Abbott:2017gyy}
including the breakthrough observation of
GW170817~\cite{TheLIGOScientific:2017qsa}, the first
combined detection of GWs and electromagnetic signals from
the same astrophysical source.
Because of the increasing sensitivity of the advanced interferometers
and planned 3rd generation detectors such as the Einstein Telescope~\cite{Punturo:2010zz}
or the Cosmic Explorer~\cite{Evans:2016mbw},
it is expected that a large number of compact binary coalescences
will be observed in the next years and decades~\cite{Aasi:2013wya,Abbott:2016nhf}.
The increasing sensitivity will not only allow
for a larger number of detections, but provide multiple signals
characterized by high signal-to-noise ratios.
These signals will be a perfect testbed for probing the nature of the
compact objects from which they arise.
Numerical relativity (NR) simulations play an essential role in the interpretation of the GW signal, elucidating the strong-field dynamics experienced during the last
stages of the binary coalescence.

Over the last decades the NR community has mainly focused on the simulation of binary systems
consisting of black holes (BHs) or neutron stars (NSs) in various combinations and initial configurations.
It is remarkable that although the breakthrough in solving the binary black hole (BBH)
problem in NR was only achieved little more than a decade ago~\cite{Pretorius:2005gq,Campanelli:2005dd,Baker:2005vv},
it has already lead to a large number of scientific discoveries such as
the simulation of the orbital hangup 
effect~\cite{Campanelli:2006uy} and showing the
existence of large recoils velocities of the final
black hole~\cite{Gonzalez:2006md,Campanelli:2007ew,Gonzalez:2007hi} or the study of
large fractions of the BBH parameter space~\cite{Mroue:2013xna,Healy:2017psd,Jani:2016wkt}.
Similarly, following the first successful NR simulations of binary neutron star (BNS)
mergers about two decades ago~\cite{Shibata:1999wm,Shibata:1999hn},
the community has made tremendous progress with an increasing coverage of the BNS parameter
space~\cite{Dietrich:2018phi}, the inclusion of microphysical aspects such as magnetic
fields~\cite{Anderson:2008zp,Giacomazzo:2010bx,Rezzolla:2011da,
Palenzuela:2013hu,Kiuchi:2014hja,Palenzuela:2015dqa,Ruiz:2017due,Kiuchi:2017zzg},
finite-temperature equation of states (EOSs), and composition effects including neutrino
transport \cite{Sekiguchi:2011zd,Galeazzi:2013mia,Neilsen:2014hha,
Sekiguchi:2015dma,Palenzuela:2015dqa,Foucart:2016rxm,Foucart:2017mbt}.

More recently, there has been a strong interest in the simulation
of exotic compact objects (ECOs) which could mimic BH or NS observations, or provide altogether new,
and as yet undetected, observational signatures.
One of the simplest potential ECOs is a boson star (BS), which is a stable solitonic solution to the coupled Einstein-Klein-Gordon equations
for a complex scalar field with gravity. The idea of a self gravitating field configuration dates back to proposals by Wheeler for ``geons"
\cite{Wheeler:1955zz}, localised self-gravitating configurations of the electromagnetic field, but was developed for complex scalar fields by Kaup
\cite{Kaup:1968zz} as well as Ruffini and Bonazzola \cite{Ruffini:1969qy}. The ideas were extended to real massive scalar fields by Seidel and Suen in
\cite{Seidel:1991zh}, with their objects later dubbed Oscillotons\footnote{Note that in the literature there is some ambiguity regarding terminology, with
BS often applied to both real and complex scalar fields. In this work we use the term BS for complex scalar field matter and
Oscillotons for real, massive scalar fields.}. Massive vector fields have also been found to support so-called Proca stars
\cite{Brito:2015yfh}. Comprehensive reviews on BSs (including the Oscilloton and Proca varieties) can be found in
~\cite{Schunck:2003kk,Liebling:2012fv}.

Overall, there has been good progress in the simulation of ECO collisions in full NR, on which we build in this work.
Most simulations have focussed on (complex scalar field) BSs,
e.g.~\cite{Palenzuela:2006wp, Choptuik:2009ww, Bezares:2017mzk, Lai:2004fw, Palenzuela:2007dm, Cardoso:2016oxy, Balakrishna:1999sv}, but other classes such as
(real scalar field) Oscillotons \cite{Helfer:2018vtq}, and just recently,
(massive, complex vector field) Proca Stars in \cite{Sanchis-Gual:2018oui,Sanchis-Gual:2017bhw} have been explored.

In the complex scalar case, the field is time-dependent - a stable, localised, oscillating configuration in the real and imaginary parts of the field
- but the spacetime geometry remains static. BSs are prevented from dispersing by the self-gravity of their time and spatial gradients and the field mass,
and are supported against gravitational collapse by the tendency of the field to disperse (which can be related to the Heisenberg Uncertainty principle).
In addition to this balance between dispersal and collapse in the free field case, self interactions of the field,
represented by a non-trivial scalar field potential $V(\phi)$, can affect the stability of the object.
In particular, adding an appropriate self interaction can provide an additional support against their collapse, allowing them to
achieve higher densities. BSs can thus be made extremely compact - comparable to NSs or even BHs.

Astrophysically, BSs represent self-gravitating objects made out of condensed bosonic particles, 
which, given a high occupation number, exist in a
macroscopic quantum state for which the expectation value may be described by a (purely classical) complex scalar field.
They are potential candidates for dark matter massive astrophysical compact halo objects 
and the cores of dark matter halos
~\cite{UrenaLopez:2010ur, Sharma:2008sc, Lee:2008ab, Lee:2008mq}
as well as providing a potential formation mechanism for the
super-massive black holes at centers of galaxies~\cite{Helfer:2016ljl}.
The only scalar boson currently known to exist is
the Higgs, but other theoretically motivated bosonic particles, such as the axion (see \cite{Marsh:2015xka} for a review), are potential candidates
for dark matter.
Models of BSs are extremely flexible. Like NSs they have a maximum stable mass which, in the free field case, scales with
the boson mass $m$, as $M_{max} \sim {M_{pl}^2}/{m}.$
Given the wide possible range of theoretical boson masses, they are capable of representing compact objects from
stellar to super-massive scales.
In addition, as noted above, changing their self interaction potential further expands the ranges
of possible masses and densities. Changing the potential in
this way can be thought of as analogous to changing the equation of state of a NS.

Whilst as yet we lack direct observational evidence for BSs, distinguishing them from more
traditional objects using GWs requires that we have a model for their GW emission. During the inspiral, this
can be partially achieved by considering the effect of the
internal structure (i.e.\ tidal effects) on the inspiral. The
tidal love numbers for different BS potentials have been
computed and simple post-Newtonian models
have been used to assess their potential distinguishability in
2nd and 3rd generation detectors~\cite{Sennett:2017etc,Johnson-McDaniel:2018uvs}.
It was found that BSs could potentially be distinguishable from NSs
with 3rd generation instruments.
For more precise predictions,
one requires a full inspiral-merger-ringdown model for binary BSs (BBSs) or
mixed object binaries.
This necessarily means using full numerical relativity to obtain information
from the strong-field regime during the merger.

Within this work we simulate what are, to the best of our knowledge, the first
configurations of mixed systems consisting of a BS with a NS companion.
For these simulations we will rely on the \BAM code~\cite{Brugmann:2008zz,Thierfelder:2011yi}
which we expand to couple the evolution of scalar fields to
the evolution of the 4-dimensional spacetime, in addition to evolving
the matter fields describing the baryonic material.
\BAM has already been employed successfully in the past for a
variety of BBH simulations, e.g.~\cite{Bruegmann:2003aw,Hannam:2007ik,Santamaria:2010yb,
Husa:2015iqa,Khan:2015jqa,Jimenez-Forteza:2016oae,Abbott:2016wiq},
BNS simulations~\cite{Dietrich:2018phi},
BH-NS systems~\cite{Dietrich:2014cea},
as well as for the study of NS
collapse~\cite{Thierfelder:2010dv,Dietrich:2014wja,Camelio:2018gfc}
and GW collapse~\cite{Hilditch:2013cba}.
Note that for now the scalar field is only coupled to the matter
sector via the gravitational interaction, i.e.,
they interact via their mutual impact on the metric,
and no additional couplings are implemented.
In the case of bosonic dark matter with a negligible coupling to
standard model physics, this approximation will be valid.
However, many scalar field candidates, such as the QCD axion,
are expected to couple to ``normal'' matter,
at least weakly, and thus their corresponding boson stars may interact with
the neutron star matter (see for example
\cite{Hook:2017psm, Eby:2017xaw, Raby:2016deh, Iwazaki:2014wka}).
Thus the exploration of such couplings,
and their impact on the merger dynamics,
represents an interesting direction for future work,
further constraining the properties of the ``dark'' sector.

The paper is organized as follow. 
We give a detailed overview of the necessary equations to be solved
for mixed BS-NS systems in Sec.~\ref{sec:equations} and discuss employed numerical
methods in Sec.~\ref{sec:methods}. We present detailed tests for single star
spacetimes in Sec.~\ref{sec:tests:single} and study BS-BS evolutions
in Sec.~\ref{sec:tests:binary}. The BS-BS mergers are compare with
results obtained from the two independent codes:
GRChombo~\cite{Clough:2015sqa} and SpEC~\cite{SpEC}.
Finally, we present mixed BS-NS configurations in Sec.~\ref{sec:BSNS}
and conclude in Sec.~\ref{sec:conclusion}.

Throughout this paper we employ geometric units, with $M_\odot=G=c=1$.
Consequently code units for lengths, times and masses are multiples of the solar mass.
To convert results stated in this paper to SI
units one has to employ the following transformations:
masses transform according to $[M] = M_\odot = 1.9889\times 10^{30}~{\rm kg}$,
lengths according to $[L]=GM_\odot/c^2\simeq1.47670\times 10^{3}$~m,
times according to $[T]=GM_\odot/c^3\simeq4.92549\times10^{-6}$~s.
Note that the geometrized boson mass $\mu$ which appears in the potential function is the quantity
$\mu = m c / \hbar$, with dimension $[L^{-1}]$ such that a value of $\mu=1$ in code units corresponds to
a particle mass of $m =1.3\times 10^{-10} ~ {\rm eV/c^2}$;
see \ref{sec:appx} for more details about unit conversion between Planck and geometric units.

\section{Governing Equations}
\label{sec:equations}

\subsection{General framework}
\label{sec:equations:general}

Following the framework of 3+1-decomposition~\cite{Arnotwitt:1960,York:1979},
see e.g.~\cite{Gourgoulhon:2007ue,Alcubierre:2008,Baumgarte:2010,Rezzolla:2013}
for textbook explanations, the metric takes the form
\begin{equation}
{\rm d}s^2  = g_{\mu \nu} {\rm d}x^\mu {\rm d}x^\nu  = -\alpha^2 {\rm d}t^2 + \gamma_{ij} ({\rm d}x^i
+ \beta^i{\rm d}t)({\rm d}x^j + \beta^j {\rm d}t),
\end{equation}
with $\alpha$ being the lapse function, $\beta^i$ the shift vector,
and $\gamma_{ij}$ the 3-dimensional spatial metric.
Here and in the following Greek indices run from $0$ to $3$
and Latin indices run from $1$ to $3$.

Decomposing Einstein's field equations
\begin{equation}
  R_{\mu \nu} - \frac{1}{2} R g_{\mu \nu} = 8 \pi T_ {\mu \nu} \label{eq:fieldeq},
\end{equation}
with the Ricci tensor $R_{\mu \nu}$, the Ricci scalar $R$, and the energy momentum tensor $T_{\mu \nu}$,
in 3+1-form leads to a set of Constraint Equations (Sec.~\ref{sec:equations:ID})
and Evolution Equations  (Sec.~\ref{sec:equations:evolution}).
While the evolution equations prescribe the time evolution of the spacetime,
the constraint equations have to remain satisfied for every time step to ensure that
the obtained solution fulfills Einstein's field equations.

To evaluate the right hand side of the field equations,
Eq.~\eqref{eq:fieldeq}, it is necessary to determine the energy momentum tensor.
We are interested in the combined simulation of bosonic and baryonic matter,
therefore,  we split the full
energy momentum tensor in a baryonic and bosonic part:
\begin{equation}
 T_{\mu \nu} = T_{\mu \nu}^{(\rho)} + T_{\mu \nu}^{(\phi)}.
\end{equation}
$T_{\mu \nu}^{(\rho)}$ corresponds to the energy-momentum tensor of the baryonic (NS) matter
and $ T_{\mu \nu}^{(\phi)}$ corresponds to the energy-momentum tensor of the bosonic matter.

Following the standard approach of general relativistic hydrodynamics (GRHD), we describe
NS material as perfect fluid with the stress-energy tensor
\begin{equation}
 T_{\mu \nu}^{(\rho)} = \rho\ h\  u_\mu u_\nu + p\  g_{\mu \nu}, \label{eq:Tmunu:NS}
\end{equation}
with the density $\rho$, enthalpy $h$, and pressure $p$,
see e.g.~\cite{Alcubierre:2008,Baumgarte:2010,Rezzolla:2013}.
The equations which determine the evolution of the matter variables turn out to be
\begin{eqnarray}
 \nabla_\mu T^{\mu \nu}  & = & 0, \label{eq:GRHD1}  \\
 \nabla_\mu (\rho u^\mu) & = & 0, \label{eq:GRHD2}  \\\
 P(\rho,\epsilon) & = & p. \label{eq:GRHD3}
\end{eqnarray}
These equations relate to the local energy-momentum conservation,
the conservation of the baryon number, and the EOS of the NS matter,
respectively.

In addition, the evolution of BSs characterized by the (complex) scalar
field $\phi$ is given by
\begin{equation}
 g^{\mu \nu} \nabla_\mu \nabla_\nu \phi = \frac{\text{d} V}{\text{d} |\phi|^2} \phi, \label{eq:KG}
\end{equation}
with the potential $V(\phi)$. The stress-energy tensor is 
\begin{align}
 T_{\mu \nu}^{(\phi)} = \frac{1}{2} \left( \nabla_\mu \phi \nabla_\nu \phi^* + \nabla_\nu \phi \nabla_\mu \phi^*\right)
                    - \frac{1}{2} \left( g^{\mu \nu} \nabla_\mu \phi \nabla_\nu \phi^* + V(|\phi|^2) \right).
\end{align}
So far, the \BAM infrastructure supports three different BS potential types.
Two choices for a complex scalar field are the free-field potential
\begin{equation}
V(\phi) = \mu^2 |\phi|^2, \label{eq:V_FF}
\end{equation}
and the solitonic potential
\begin{equation}
V(\phi) = \mu^2 |\phi|^2 \left(1 - 2 \frac{|\phi|^2}{\sigma^2} \right)^2, \label{eq:V_Sol}
\end{equation}
where $\mu$ relates to the geometrised mass of the considered
bosons (see \ref{sec:appx}) and $\sigma$ determines the compactness of the solitonic BS.
Furthermore, as discussed in detail in~\cite{Clough:inprep}, the current infrastructure
also permits the simulation of real scalar fields in combination with the free-field
potential \eqref{eq:V_FF} or the axion potential
\begin{equation}
V(\phi) =  2 f_a^2 \mu^2 \left(1 - \cos\left(\frac{\phi}{f_a}\right)\right), \label{eq:V_axion}
\end{equation}
with $f_a$ being the axion decay constant.

\subsection{Matter-source terms}

To compute the right hand side of Einstein's field equations we have to derive the
standard York-ADM matter variables, i.e., the energy density ($E$),
the momentum density $(S^i)$, the trace of the momentum density ($S$),
and the spatial part of the energy-momentum tensor ($S_{ij}$).
Those quantities are generally given by
\begin{eqnarray}
  E      & = & \ \ T_{\mu \nu} n^\nu n^\mu, \label{eq:defE} \\
  S^i    & = & - T_{\mu \nu} n^\mu \gamma^{\nu i}, \label{eq:defSi} \\
  S^{ij} & = & \ \ T_{\mu \nu} \gamma^{\mu i } \gamma^{\nu j}, \label{eq:defSij} \\
  S      & = & \gamma^{ij}S_{ij} \label{eq:defS},
  \end{eqnarray}
where for the baryonic NS matter, we can write the ADM matter variables explicitly as
\begin{eqnarray}
E^{(\rho)}      & = & \rho h W^2 - p, \\
S^{(\rho)}      & = & \rho h W^2  v^i v_i +3 p, \\
S_{ij}^{(\rho)} & = & \rho h W^2 v_i v_j + \gamma_{ij} p, \\
S^i_{(\rho)}   & = & \rho h W^2  v^i,
\end{eqnarray}
with the Lorentz factor $W=1/\sqrt{1-v^2}$ and $v^i$ being the fluid
velocity by a Lagrangian observer.
The ADM matter variables for the bosonic matter are
\begin{eqnarray}
E^{(\phi)}      & = & \frac{    |\Phi|^2 + |\Pi|^2  + V    }{2}, \\
S^{(\phi)}      & = & \frac{  3 |\Pi|^2  - |\Phi|^2 - 3 V  }{2},\\
S_{ij}^{(\phi)} & = & \frac{  (\Phi_i \Phi_j^*+\Phi_j \Phi_i^*) - \gamma_{ij} (|\Phi|^2-|\Pi|^2+V)}{2}, \\
S^i_{(\phi)}    & = & \gamma^{ij} \frac{  \Pi \Phi^*_j + \Pi^* \Phi_j }{2},
\end{eqnarray}
with $\Phi_i = \partial_i \phi, |\Pi|^2 = \Pi \Pi^*, |\Phi|^2 = \gamma^{ij} \Phi_i \Phi_j^*$.

\subsection{Initial configurations}
\label{sec:equations:ID}

The current procedure for the construction of boson-baryon star configurations is as follows:
\begin{enumerate}[(i)]
 \item We compute the metric and matter fields for isolated NSs or BSs.  
       In cases for orbiting configurations we additionally boost the individual objects.
 \item We superimpose the individual spacetimes to obtain a first initial guess for the binary evolution.
 \item We solve the constraint equations to obtain a solution consistent with general relativity.
\end{enumerate}

This methods allows for the construction of generic systems, but has a significant drawback,
namely the possibility that the stars get artificially excited. The reason for this is that
we do not re-solve for the matter fields with the adjusted spacetime
background computed in step (iii). Therefore, the simulated NSs and BSs are not
necessarily in their ground state and show additional oscillations.
While there has been progress in constructing
NSs in hydrodynamical equilibrium for generic
configurations~\cite{Tichy:2011gw,Kyutoku:2014yba,
Moldenhauer:2014yaa,Dietrich:2015pxa,Tacik:2015tja},
no equivalent formalism exists, to the best of our knowledge,
for BSs.

\subsubsection{Constructing isolated boson stars:}

To obtain the stable solution for spherically symmetric boson stars we start from the harmonic ansatz for the
scalar field
\begin{equation}
\phi=\phi_{0}e^{i\omega t}
\end{equation}
where $\omega$ is a constant oscillation frequency of the field and $\phi_{0}(r)$ is
a real-valued spatial profile.
We assume that the metric is static and use maximal slicing so that the extrinsic
curvature vanishes, $K\equiv 0$. Under these assumptions, the metric can be written as
\begin{equation}
{\rm d}s^2= -\alpha^{2} {\rm d}t^{2} + a^{2}{\rm d}r^{2} + r^{2}{\rm d}\Omega^{2}.
\end{equation}

Given our assumptions,
the momentum constraint is satisfied identically,
while the Hamiltonian constraint,
the slicing condition, and the Klein-Gordon equation
yield the following system of equations:
\begin{align}
\partial_ra &= -\frac{a}{2r}(a^{2}-1)+4\pi r a^{3}E^{(\phi)}, \\
\partial_r\alpha &= \frac{\alpha}{2r}(a^{2}-1)+4\pi r\alpha a^{2}S^{(\phi)}, \\
 \partial_r\Phi_0 &=-\left[1+a^{2}+4\pi r^{2}a^{2}(S^{(\phi)}-E^{(\phi)})\right]\frac{\Phi_0}{r}-
 \left(\frac{\omega^{2}} {\alpha^{2}}-V'\right)\phi_0a^{2}, \\
 \partial_{r}\phi_{0} &= \Phi_{0}.
\end{align}
The boundary conditions at the origin are fixed by
demanding smoothness and
the boundary conditions at infinity come from imposing asymptotic flatness, i.e.,
\begin{align}
	\phi_{0}(0) \equiv \phi_{c}, \qquad   & \qquad
	\Phi_{0}(0)  = 0, \\
	\lim_{r\rightarrow\infty} a  = 1, \qquad   & \qquad
	\lim_{r\rightarrow\infty} \alpha  = 1.
\end{align}
We set the boundary conditions at infinity by using a compactified radial coordinate.

The resulting system of ordinary differential equations (ODEs)
is solved by a collocation method {\tt bvpcol}
from the {\tt bvpSolve} R package\cite{Mazzia2014}\footnote{\url{https://cran.r-project.org/package=bvpSolve}}.
In many cases, especially for solitonic BSs,
a good initial guess is necessary to obtain convergence.
We obtain solutions at different values of the central field
by slowly varying it and using previous solutions as starting points.
We transform the solution from
areal polar coordinates to isotropic coordinates and interpolate this onto the numerical grid.

To obtain a boosted configuration (for which we assume the boost along the $y$ direction) 
we follow the method outlined
in~\cite{Bezares:2017mzk}. The scalar field is then given by: 

\begin{eqnarray} 
 \mathcal{R}(\phi)  = |\phi_0| \cos (\xi)\\
 \mathcal{I}(\phi) = - |\phi_0| \sin (\xi)\\ 
 \mathcal{R}(\Pi)  = \frac{\Gamma \omega |\phi_0| (1+\beta^y \mathcal{P})
                      \sin (\xi)}{\alpha}+
               \frac{\Gamma^2y (\beta^y + \mathcal{P}) \partial_r|\phi_0| \cos(\xi)} 
               {(\alpha r/B_0)} \\
 \mathcal{I}(\Pi)  = \frac{\Gamma \omega |\phi_0| (1+\beta^y \mathcal{P})
                      \cos (\xi)}{\alpha}-
               \frac{\Gamma^2y (\beta^y + \mathcal{P}) \partial_r|\phi_0| \sin(\xi)} 
               {(\alpha r/B_0)} \\               
\end{eqnarray}
with $\xi = \theta-\Gamma \mathcal{P} y \omega$, 
$B_0=\Gamma \sqrt{1-(\mathcal{P} \alpha)^2/\psi^2}$, 
$\Gamma = \sqrt{1-\mathcal{P}^2}$. Furthermore $\phi_0$ denotes the scalar 
field of the unboosted star, $\mathcal{P}$ is the boosting parameter, and $\theta$
a possible shift in the scalar field's phase. 

Note that after changing the scalar field profile, we also adjust the spacetime 
background by following the procedure outlined in Sec.~\ref{sec:equations:ID:CTS}.

\subsubsection{Constructing isolated neutron stars:}

We obtain isolated, spherically symmetric NSs by
solving the TOV equation~\cite{Tolman:1939,Oppenheimer:1939}.
For this purpose we write the four-metric in areal polar coordinates as
\begin{equation}
 {\rm d}s^2= - e^{2 \nu} {\rm d}t^2 + \left( 1- \frac{2 m}{R} \right)^{-1} {\rm d}R^2+R^2 {\rm d}\Omega^2.
\end{equation}
To obtain $m(R),\nu(R)$, and the pressure $p(R)$, the TOV equations
\begin{eqnarray}
 \frac{{\rm d} \rho}{{\rm d} R}& = & \left(\rho (1+\epsilon)+p\right)
 \frac{m+4 \pi r^3 p}{R (R-2m)}\cdot \frac{1}{\frac{{\rm d}p} {{\rm d} \rho}},  \\
 \frac{{\rm d} m}{{\rm d} R}& = & 4 \pi R^2 \rho (1+\epsilon),  \\
 \frac{{\rm d} \nu}{{\rm d} R}& = & \frac{ m + 4 \pi R^3 p}{R(R-2m)},
\end{eqnarray}
are solved with an explicit fourth order Runge-Kutta algorithm.
As initial conditions for $R=0$,
we choose $\rho(R=0)=\rho_{\rm c}, m(R=0)=0,$ and  $\nu(R=0)=0$.
Once the $\rho(R),m(R),\nu(R)$ are known, we transform the solution from
areal polar coordinates to isotropic coordinates.

The employed TOV solver is part of the \BAM code,
see~\cite{dbt_mods_00019798,Moldenhauer:2014yaa,Bugner:2015gqa}
for more details.

\subsubsection{Conformal Thin Sandwich Equations:}
\label{sec:equations:ID:CTS}

Following the 3+1-framework, the Einstein constraint equations are given as
\begin{eqnarray}
{^{(3)}}R+K^2-K_{\alpha \beta} K^{\alpha \beta} & = & 16 \pi E , \label{eq:admham}\\
D_jK^{i j} - D^i K& = & 8 \pi S^i , \label{eq:admmom}
\end{eqnarray}
with the extrinsic curvature $K_{ij}$.
Eq.~\eqref{eq:admham} refers to the Hamiltonian constraint and
Eqs.~\eqref{eq:admmom} are the three momentum constraints.
$D_i$ denotes the three-dimensional covariant derivative
after projecting the standard covariant derivative onto
the space orthogonal to normal vector of the hypersurface.

To simplify the constraint equation system, we split the three-metric into a
conformal factor $\psi$ and the corresponding conformal metric
$\bar{\gamma}_{ij}$, writing
\begin{equation}
  \gamma_{ij} = \psi^4 \bar{\gamma}_{ij}. \label{eq:confdecomp1}
\end{equation}
Similarly, we express the extrinsic curvature in terms of a
trace-free piece $A_{ij}$, writing
\begin{equation}
  K_{ij} = A_{ij} +\frac{1}{3}\gamma_{ij} K.  \label{eq:confdecomp2}
\end{equation}
Assuming again maximal slicing ($K=0$) and setting $\partial_t \bar{\gamma}_{ij} \gamma_{ij} =0$ as well as
\begin{equation}
  A^{ij} = \frac{1}{2\psi^4 \alpha} (\textbf{L} \beta)^{ij} ,
\end{equation}
we obtain with~\eqref{eq:confdecomp1} and~\eqref{eq:confdecomp2} the constraint equations
within the conformal thin-sandwich (CTS)
approach, e.g.~\cite{Wilson:1995uh,Wilson:1996ty,York:1998hy}:
\begin{align}
  \bar{D}^2 \psi & = - \frac{\psi^5}{32 \alpha^2} (\textbf{L} \beta)^{ij} (\textbf{L} \beta)_{ij} - 2 \pi \psi^5 E, \label{eq:metric1} \\
  \bar{D}_j (\textbf{L} B)^{ij} &= (\textbf{L} B)^{ij} \bar{D}_j
  \ln\left(\frac{\alpha}{\psi^6}\right) + 16 \pi \alpha \psi^4 S^i, \label{eq:metric2} \\
  \bar{D}^2(\alpha \psi) & = \alpha \psi
  \left( \frac{7 \psi^4}{32 \alpha^2} (\textbf{L} \beta)^{ij} (\textbf{L} \beta)_{ij} + 2 \pi \psi^4 (E + 2S)
  \right), \label{eq:psialpha}
\end{align}
with $(\textbf{L} \beta)^{ij}= \bar{D}^i \beta^j + \bar{D}^j \beta^i
- \frac{2}{3} \delta^{ij} \bar{D}_k \beta^k$. $\bar{D}_i$
denotes the flat-space covariant derivative in Cartesian
coordinates, i.e.~$\bar{D}_i =\partial_i$.

Note that during the construction of CTS initial data, 
we found that a fixed NS density profile leads to an overall larger baryonic mass
due to solving Eq.~\eqref{eq:metric1}. Consequently,
the solved solution is not identical to the setup we are looking for. 
%this is not the setup you are looking for...<waves hand>
In some cases this can even lead to unstable solutions in which the 
NS collapses immediately to a BH. 
To avoid this effect there are two simple solutions - either 
the initial separation is increased (which is possible for head-on collisions but 
not computationally affordable for orbiting binaries)
or the density profile may be adjusted during the iteration procedure.
To adjust the density profile we adjust the density according to
\begin{equation}
\rho_{\rm new} = \psi_{\rm old}^4 \ \frac{\rho_{\rm old}}{\psi_{\rm new}^4} . \label{eq:rho_update}
\end{equation}
According to the adjusted density profile we also reconstruct 
the enthalpy according to the given EOS, Eq.~\eqref{eq:GRHD3}.

An iterative procedure of solving the CTS equations and
updating the density via Eq.~\eqref{eq:rho_update}
leads to the desired configuration. 
Note further that during the iterative procedure the NS's velocity is given by 
\begin{equation}
 v^i = (\beta^x/\alpha, \beta^y/\alpha+\mathcal{P},\beta^z/\alpha), 
\end{equation}
assuming again a boost with parameter $\mathcal{P}$ along the $y$ axis.

\subsection{3+1-Evolution Equations}
\label{sec:equations:evolution}

\subsubsection{Spacetime evolution:}

For the spacetime evolution \BAM employs the
BSSNOK~\cite{Nakamura:1987zz,Shibata:1995we,Baumgarte:1998te} or the
Z4c scheme~\cite{Bernuzzi:2009ex,Hilditch:2012fp}.
Both schemes have been tested in the past for vacuum and
NS spacetime evolutions.
Due to the improved constraint propagation and
damping properties of Z4c, we use this evolution scheme
as our preferred choice. We recast it in the following for completeness.

In a similar way to the construction of the initial configurations of
the conformal metric and conformally rescaled extrinsic curvature, we introduce:
 \begin{eqnarray}
 \tilde{\gamma}_{ij} & = & \chi \gamma_{ij} \label{eq:Z4-1},\\
 \tilde{A}_{ij} &=& \chi (K_{ij} -\frac{1}{3} \gamma_{ij} K) \label{eq:Z4-2},\\
 \hat{K} & = & \gamma^{ij} K_{ij} - 2 \Theta, \label{eq:Z4-3}
 \end{eqnarray}
with the constraint $\Theta$.

The full Z4c evolution equation systems is then given by
\begin{small}
 \begin{eqnarray}
 \partial_t \chi & = & \frac{2}{3} \chi \left( \alpha ( \hat{K} + 2 \Theta) - D_i \beta^i \right), \\
 \partial_t \tilde{\gamma}_{ij} & = & - 2 \alpha \tilde{A}_{ij} + \beta^k  \partial_k \tilde{\gamma}_{ij} + 2 \tilde{\gamma}_{k(i} \partial_{j)} \beta^k  - \frac{2}{3} \tilde{\gamma}_{ij} \partial_k \beta^k,\\
 \partial_t \hat{K} & = & - D^i D_i \alpha + \alpha \left( \tilde{A}_{ij} \tilde{A}^{ij} + \frac{1}{3} ( \hat{K} + 2 \Theta) ^2 \right) \nonumber \\
                        && + 4 \pi \alpha (S + E) + \beta^k \partial_k \hat{K} + \alpha \kappa_1 (1-\kappa_2) \Theta, \\
 \partial_t \tilde{A}_{ij} & = & \chi \left( -D_i D_j \alpha + \alpha \ ({^{(3)} R_{ij}}- 8 \pi S_{ij} ) \right)^{TF}
                               + \alpha \left( ( \hat{K} + 2 \Theta) \tilde{A}_{ij} - 2 {{\tilde{A}^k}}_{\ j} \tilde{A}_{kj} \right) \nonumber \\
                              && + \beta^k \partial_k \tilde{A}_{ij} + 2 \tilde{A}_{k(i} \partial_{j)} \beta^k- \frac{2}{3} \tilde{A}_{ij} \partial_k  \beta^k, \\
 \partial_t \tilde{\Gamma}^i & = &
 -2\tilde{A}^{ik} \partial_k \alpha + 2 \alpha \left( \tilde{\Gamma}^i_{kl} \tilde{A}^{kl} - \frac{3}{2} \tilde{A}^{ik} \partial_k \ln(\chi)
 -\frac{1}{3} \tilde{\gamma}^{ik} \partial_k (\hat{K} + 2 \Theta) - 8 \pi \tilde{\gamma}^{ik} S_k \right) \nonumber \\
 && + \tilde{\gamma}^{kl} \partial_k \partial_l \beta^i +\frac{1}{3} \tilde{\gamma}^{ik} \partial_l \partial_k \beta^l - 2 \alpha \kappa_1 (\tilde{\Gamma}^i - \bar{\Gamma}^i)
 + \beta^k \partial_k \tilde{\Gamma}^i  \nonumber \\ && - \bar{\Gamma}^k \partial_k \beta^i + \frac{2}{3} \bar{\Gamma}^i \partial_k \beta^k, \\
 \partial_t \Theta &= & \frac{\alpha}{2} \left( {^{(3)} R} - \tilde{A}_{ij} \tilde{A}^{ij} + \frac{2}{3} (\hat{K} + 2 \Theta)^2 \right)
 - \alpha \left( 8 \pi E + \kappa_1 (2+\kappa_2) \Theta \right) \nonumber \\ & & + \beta^i \partial_i \Theta ,
 \end{eqnarray}
 \end{small}
with $\tilde{\Gamma}^i = 2 \tilde{\gamma}^{ik} Z_k + \tilde{\gamma}^{ij} \tilde{\gamma}^{kl} \tilde{\gamma}_{jk,l}$ and
 $\bar{\Gamma}^i = \tilde{\gamma}^{kl} \tilde{\Gamma}^i_{,kl}$.
 The two parameters $\kappa_1$ and $\kappa_2$ are the Z4c-damping parameters and will be studied in more detail
 during single star spacetime evolutions in Sec.~\ref{sec:tests:single}.

\subsubsection{Scalar field evolution:}

The Klein-Gordon equation, Eq.~\eqref{eq:KG}, describes the evolution of
the scalar field governing the bosonic matter.
Introducing the conjugate momenta of the
scalar field $\Pi = \mathcal{L}_n \phi$, the Klein-Gordon equation can be written in 3+1-form as
\begin{eqnarray}
 \partial_t \phi & = &  \beta^k \partial_k \phi - \alpha \Pi, \label{eq:evo:phi} \\
 \partial_t \Pi &= & \beta^k \partial_k \Pi + \alpha \left[- \chi \tilde{\gamma}^{ij} \partial_i \partial_j \phi +
 \chi \Gamma^i \partial_i \phi + \frac{1}{2} \tilde{\gamma}^{ij} \partial_i \phi \partial_j \chi \right. \nonumber \\
 && \left. + \Pi \hat{K} + \frac{\text{d}V}{\text{d}|\phi|^2} \phi \right] -
 \chi \tilde{\gamma}^{ij} \partial_i \phi \partial_j \alpha. \label{eq:evo:pi}
\end{eqnarray}
Since $\phi$ and $\Pi$ can be split into real and imaginary parts,
Eqs.~\eqref{eq:evo:phi} and \eqref{eq:evo:pi} are split into 4 equations
for the evolution of $\R(\phi),\I(\phi), \R(\Pi), \I (\Pi)$.

\subsubsection{General Relativistic Hydrodynamics:}

For the simulation of the baryonic matter, we
follow~\cite{Banyuls:1997zz}. A detailed discussion about the \BAM implementation is given
in~\cite{dbt_mods_00019798,Thierfelder:2011yi}. The main aspects will be reported here for completeness,
but have not been modified for the scope of this paper with respect to previous \BAM versions.
To describe the equations of general relativistic hydrodynamics (GRHD)
a set of primitive variables $\textbf{w}=(\rho,v_i,\epsilon)$,
(rest-mass density, the fluid velocity, and the internal energy density measured
by a Lagrangian observer, respectively) and a set of conservative variables
$\mathbf{q}=(D,S_i,E)$ (the conserved rest mass, momentum, and internal energy density
of the Eulerian observer, respectively) is introduced.
The primitive and conservative variables are connected by
\begin{subequations}
\begin{eqnarray}
    D & = & \rho W \label{eq:consv1} \\
    S_i & = & \rho h W^2 v_i \label{eq:consv2}\\
    \tau & = & \rho h W^2 - p -D \label{eq:consv3}
\end{eqnarray}
\label{eq:consvall}
\end{subequations}
with the Lorentz factor $ W = 1/\sqrt{1 - \gamma_{kl} v^k v^l}$.
The enthalpy, used to define the energy-momentum tensor Eq.~\eqref{eq:Tmunu:NS},
is related to the primitive variables via $h=1+\epsilon+p/\rho$.
Rewriting Eqs.~\eqref{eq:GRHD1}-\eqref{eq:GRHD3} as first-order,
flux conservative, hyperbolic system gives
\begin{equation}
   \frac{1}{\sqrt{-g}} \left( \frac{\partial (\sqrt{\gamma} \textbf{q})}{\partial x^0}
 + \frac{\partial (\sqrt{-g} \textbf{F}^i ) }{\partial x^i} \right) = \textbf{S,} \label{eq:mattereq}
\end{equation}
where
\begin{eqnarray}
\label{eq:num_methods:mattereqU}
\textbf{q}(\textbf{w}) & = &  (D, S_j, \tau),\\
\label{eq:num_methods:mattereqF}
\textbf{F}^i(\textbf{w}) & = & \left( D \left(v^i- \frac{\beta^i}{\alpha}\right),
                  S_j\left(v^i-\frac{\beta^i}{\alpha} \right)+ p \delta^i_j,\tau \left( v^i-\frac{\beta^i}{\alpha} \right) + p v^i \right), \\
\label{eq:num_methods:mattereqS}
\textbf{S}(\textbf{w}) &
             = & \left( 0, T^{\mu \nu} \left( \partial_\mu g_{\nu j} - \Gamma^\sigma_{\nu \mu} g_{\sigma j} \right), \alpha \left(  T^{\mu 0} \partial_\mu (\log(\alpha)) -T^{\mu \nu} \Gamma^0_{\nu \mu} \right) \right).
\end{eqnarray}

\section{Numerical Methods}
\label{sec:methods}

\subsection{Initial data construction}

Following the discussion in Sec.~\ref{sec:equations:ID},
we obtain an initial guess for the metric
and matter fields based on a superposition of the isolated BS
and/or NS solutions.
Afterwards, we ensure that the described system is a solution of GR
by solving the constraint equations in CTS form.

To solve the set of coupled elliptic equations,
we employ \BAM's multigrid
solver as described in detail in~\cite{Moldenhauer:2014yaa}.
The multigrid solver uses nested boxes on a Cartesian grid and
approximates derivatives in each substep of the iteration procedure
by standard finite difference stencils
(here second order stencils are applied).

\subsection{Dynamical evolutions}

The evolution scheme of the \BAM code is based on the method of lines.
\BAM uses an adaptive mesh refinement (AMR) employing the method of 'moving boxes' in which
the domain consists of a hierarchy of nested Cartesian grids (refinement levels).
Each finer refinement level has half the grid spacing of its surrounding coarser level.
Innermost levels move dynamically during the time evolution
following the motion of the stars. This makes it possible to cover the strong field region
with the highest resolution, but also extract GWs or ejecta quantities sufficiently far away from the
compact binary system. \BAM also has the capability to add a ``cubed-sphere''
multi-patch AMR for the coarsest level to allow a more
accurate extraction of GWs~\cite{Hilditch:2012fp}.

Time integration is performed following the Berger-Oliger~\cite{Berger:1984zza,Brugmann:2008zz}
or Berger-Collela~\cite{Berger:1989a,Dietrich:2015iva}
method allowing sub-cycling in time for different refinement levels to reduce computational costs.
Each individual refinement level is typically evolved with a 4th order Runge-Kutta algorithm.
The impact of the time integrator and Courant factors
is discussed in more detail in the next section.

Derivatives of spacetime variables and bosonic fields are computed using
4th or 6th order finite differences, see Sec.~\ref{sec:tests:single}.
Additionally, 6th or 8th order artificial dissipation is added to stabilize noise from mesh
refinement boundaries.

GRHD equations were solved with a second order scheme based on the local Lax-Friedrich
scheme for the flux computation, and primitive reconstruction.
The flux reconstruction uses WENOZ~\cite{Borges:2008a,Bernuzzi:2012ci} 
a fifth-order Weighted Essentially Non-Oscillatory scheme. 
We do not employ high-order methods as discussed in~\cite{Bernuzzi:2016pie}
for these first tests of BSNS setups.
For our simulations vacuum regions are described by a static, low-density, and cold
atmosphere~\cite{Thierfelder:2011yi}.
The atmosphere densities are typically $\sim 12$ orders of magnitude below the central density of the NSs we are simulating, however,
low-density flow is one of the main error sources for the NS simulations,
\cite{Radice:2013xpa,Bernuzzi:2016pie,Guercilena:2016fdl}.

\section{Testbeds: Single star spacetimes}
\label{sec:tests:single}

\begin{figure}[t]
\centering
\includegraphics[width=0.95\textwidth]{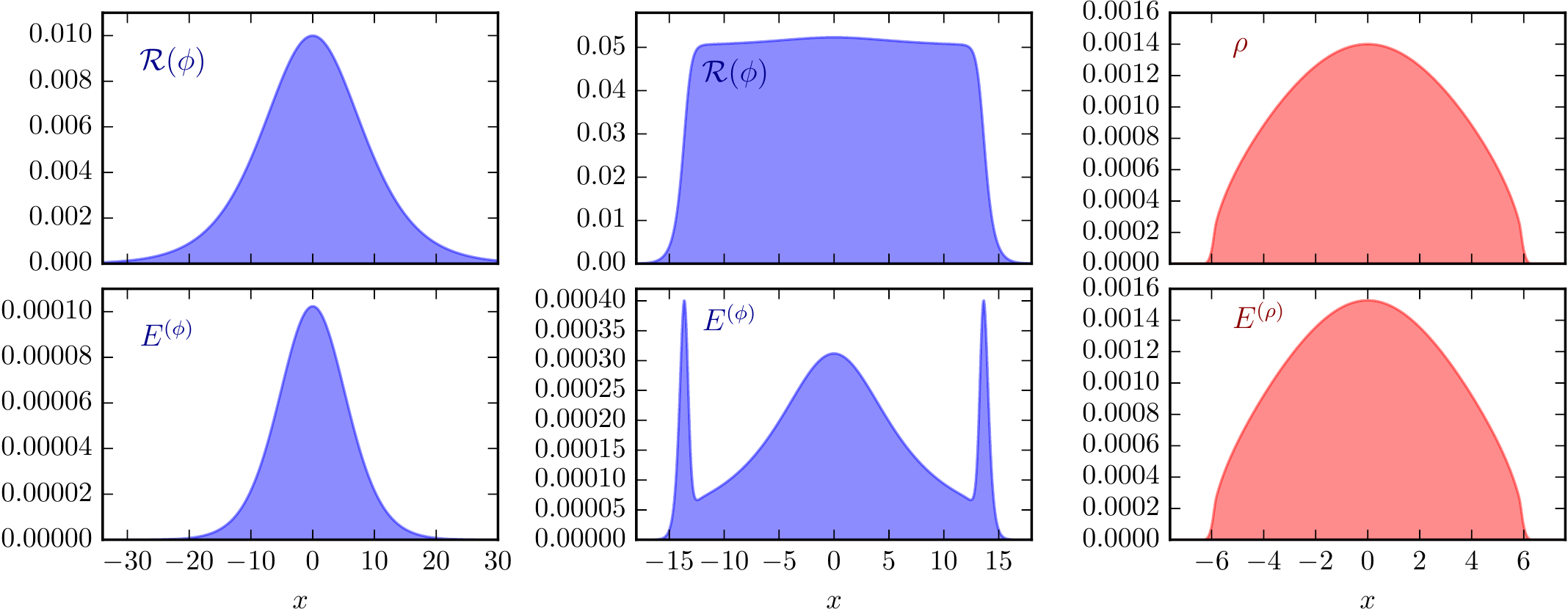}
\caption{Star profiles [real part of the scalar field for BSs (blue) and density for NSs (red)]
and energy density for the test cases studied in this section, Tab.~\ref{tab:tests-grid}.
From left to right:
the free-field BS with a mass of $M=0.36$,
the solitonic BS with a mass of $M=7.35$,
and a NS employing the SLy EOS with mass $M=1.35$ are presented.}
\label{fig:profile_single_stars}
\end{figure}

As a first testbed,
we perform single star evolutions of (i) a free-field BS,
(ii) a solitonic BS, and (iii) an isolated NS employing the SLy
EOS~\cite{Douchin:2001sv,Read:2008iy,Dietrich:2015pxa}.
We will focus our discussion mostly on the simulation of the isolated BSs,
but present some NSs evolutions for completeness. We refer
to~\cite{Thierfelder:2011yi,Dietrich:2015iva,Bernuzzi:2016pie}
for detailed studies of TOV simulations with \BAM.

For illustrative purposes, we show the profile of the
real part of the scalar field [$\mathcal{R}(\phi)$]
for the BSs and the density [$\rho$] for the studied NS in
the top panel of Fig.~\ref{fig:profile_single_stars}.
The bottom panel of Fig.~\ref{fig:profile_single_stars}
shows the corresponding energy density.
It is evident that:
(i) NSs show a clear surface while such a surface is absent for BSs;
(ii) solitonic BSs show a peak in the energy density close to the star's surface;
(iii) the solitonic potential allows for more compact stars keeping the 
geometrized boson mass $\mu$ fixed.

\begin{table}[t]
  \centering
  \caption{Overview of single star evolutions.
    The columns refer to: 
    the simulations name,
    the employed potential/EOS,
    the gravitational mass of the star,
    the total number of boxes ($L$),
    the finest non-moving level ($l^{mv}$),
    the number of points in the fixed (moving) boxes ($n$ $(n^{mv})$),
    the grid spacing in level $l=0,L-1$ ($h_{0},h_{L-1}$).
    The resolution in level $l$ is $h_l=h_0/2^l$. }
  \begin{tabular}{l|cccc|cccc}
     name                            & potential/EOS  & M  & $L$ & $l^{mv}$ & $n$ & $n^{mv}$ & $h_0$ & $h_{L-1}$  \\
     \hline
     BS$_{\rm static}^{\rm FF}$-R1   & free-field     & $0.361$     & 5   & 2        & 80  & 48       &  4.0  & 0.250  \\
     BS$_{\rm static}^{\rm FF}$-R2   & free-field     & $0.361$     & 5   & 2        & 160 & 96       &  2.0  & 0.125  \\
     BS$_{\rm static}^{\rm FF}$-R3   & free-field     & $0.361$     & 5   & 2        & 320 & 192      &  1.0  & 0.0625 \\
     \hline
     BS$_{\rm static}^{\rm Sol}$-R1  & solitonic   &  $7.346$ & 5   & 2        & 160 & 160      &  4.0  & 0.250  \\
     BS$_{\rm static}^{\rm Sol}$-R2  & solitonic   &  $7.346$ & 5   & 2        & 320 & 320      &  2.0  & 0.125  \\
     BS$_{\rm static}^{\rm Sol}$-R3  & solitonic   &  $7.346$ & 5   & 2        & 640 & 640      &  1.0  & 0.0625 \\
     \hline
     NS$_{\rm static}^{\rm SLy}$-R1  & SLy         &  $1.350$ & 5   & 4        & 64  & 64       &  3.6  & 0.225  \\
     NS$_{\rm static}^{\rm SLy}$-R2  & SLy         &  $1.350$ & 5   & 4        & 128 & 128      &  1.8  & 0.113  \\
     NS$_{\rm static}^{\rm SLy}$-R3  & SLy         &  $1.350$ & 5   & 4        & 256 & 256      &  0.9  & 0.056  \\
     \hline
     BS$_{\rm boost}^{\rm FF}$-R1    & free-field  & $0.361$     & 4   & 1        & 160 & 48       &  8.0  & 1.0   \\
     BS$_{\rm boost}^{\rm FF}$-R2    & free-field  & $0.361$     & 4   & 1        & 320 & 96       &  4.0  & 0.5   \\
     BS$_{\rm boost}^{\rm FF}$-R3    & free-field  & $0.361$     & 4   & 1        & 640 & 192      &  2.0  & 0.25  \\
  \end{tabular}
 \label{tab:tests-grid}
\end{table}

\subsection{Stationary boson stars: free-field potential}

\begin{figure}[t]
\centering
\includegraphics[width=1.\textwidth]{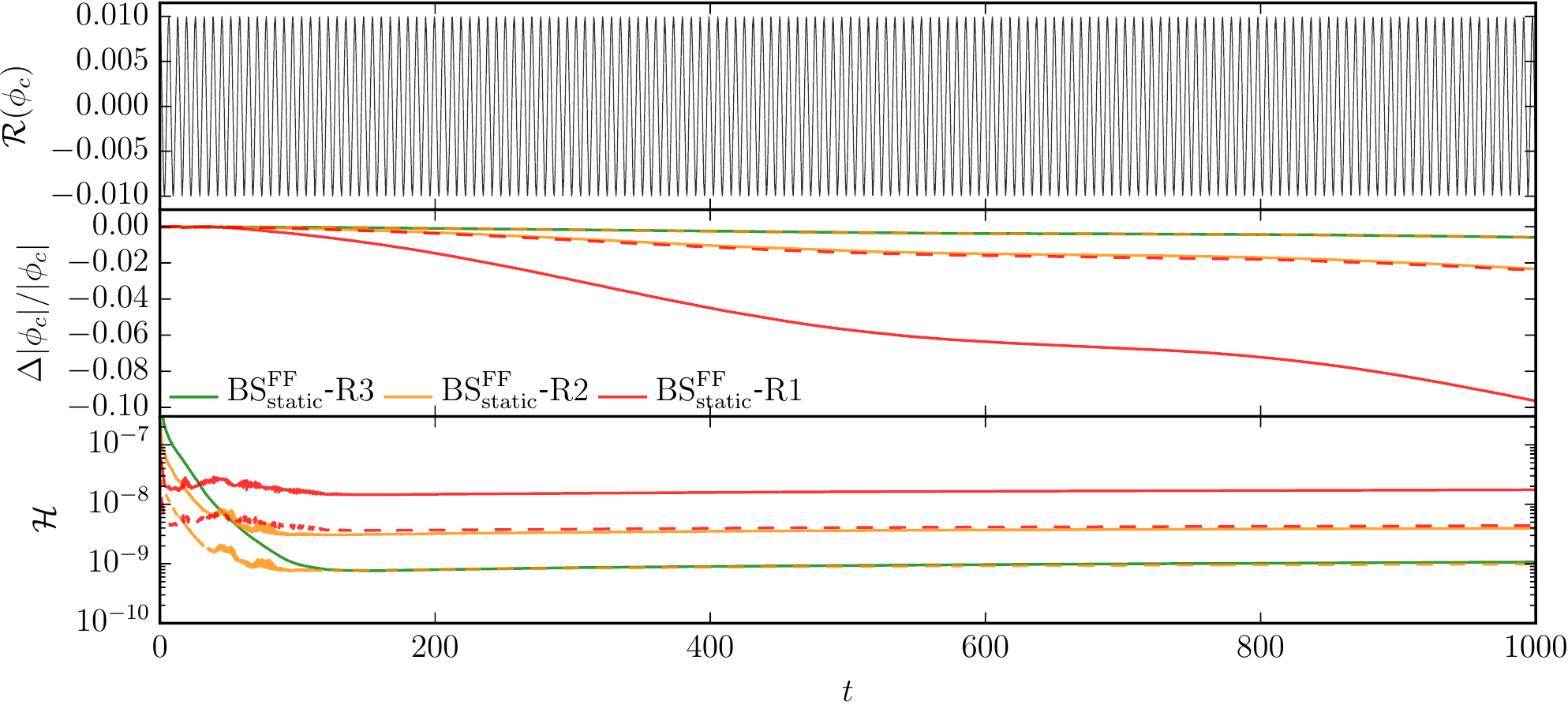}
\caption{Effect of resolution on the isolated BS evolution for free-field potential.
The panels show the central value of the real part of the scalar field $\phi_c$,
the relative difference of the scalar field amplitude at the center of the star,
and the Hamiltonian constraint. Different colors refer to different resolutions,
Tab.~\ref{tab:tests-grid}. Dashed lines show rescaled results assuming second order convergence.}
\label{fig:single_BS_FF_res}
\end{figure}

\begin{figure}[t]
\centering
\includegraphics[width=1.\textwidth]{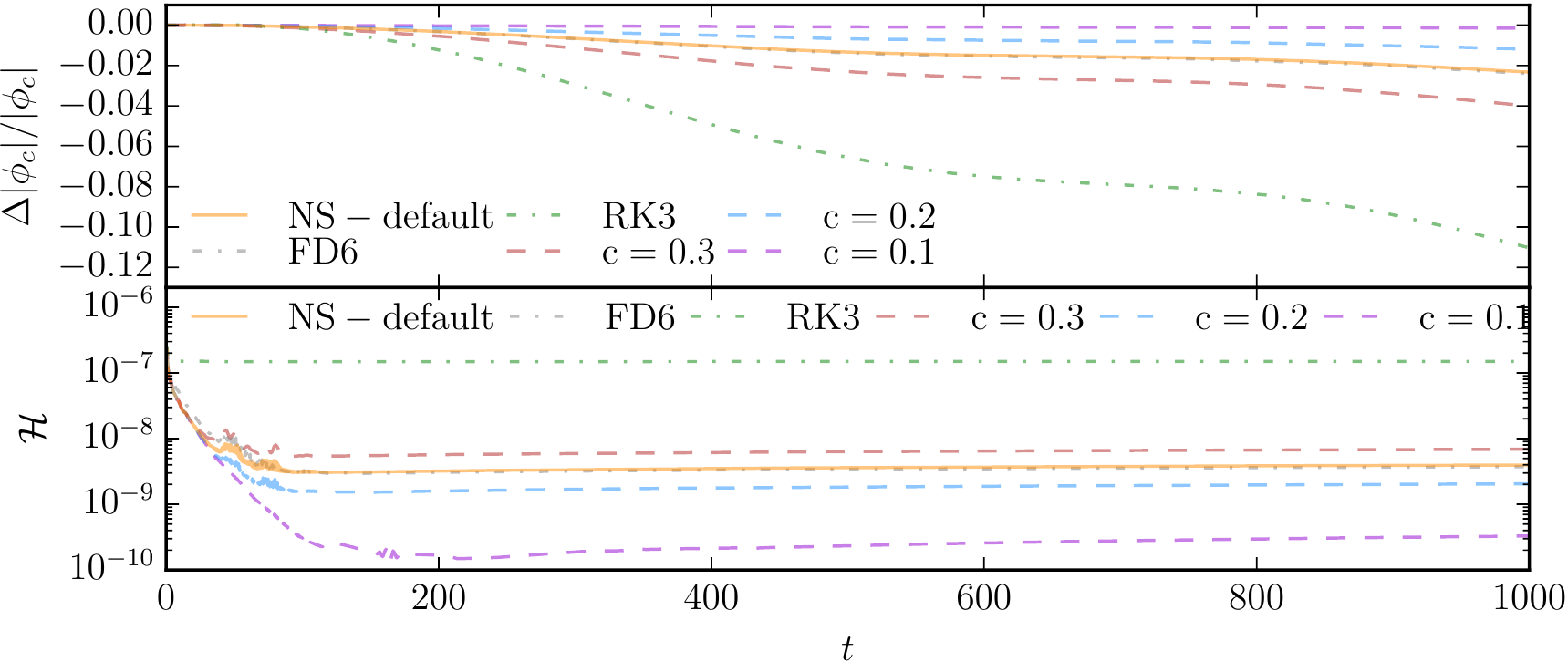}
\caption{Effect of the time integrator, CFL condition, and order of the spatial derivatives
on the central amplitude of the scalar field (top) and Hamiltonian constraint (bottom).
As a default setup, we use a 4th order Runge-Kutta scheme with a CFL condition
of $c=0.25$ with 4th order finite differencing stencils for the spatial derivatives,
we refer to this as 'NS-default', since it refers to our default choice for the simulation
of NS spacetimes. We employ the $\rm R2$ resolution.}
\label{fig:single_BS_FF_RKcfl}
\end{figure}

\begin{figure}[t]
\centering
\includegraphics[width=1.\textwidth]{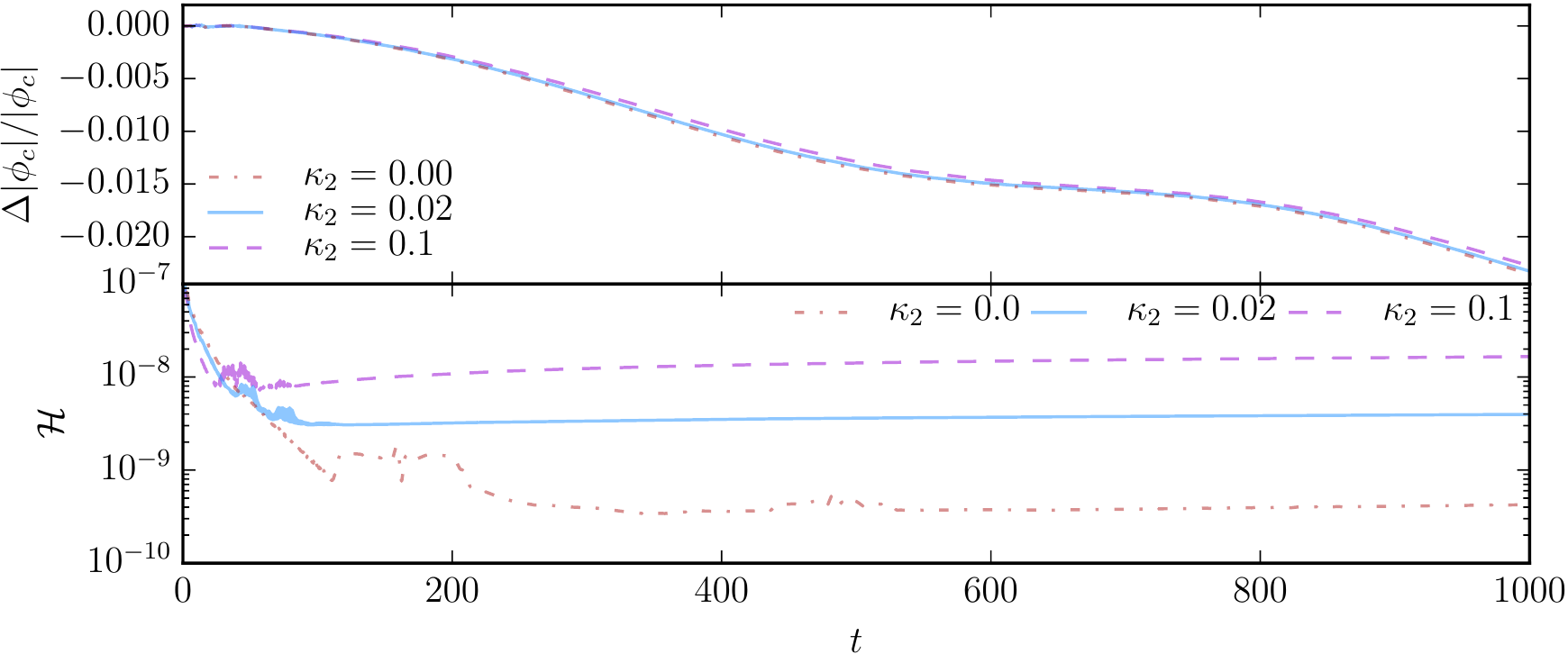}
\caption{Influence of the $\kappa_1$ damping parameter on the central value of $|\phi|$ (top panel)
and on the Hamiltonian constraint (bottom panel). For all tests we set $\kappa_1=0$.}
\label{fig:single_BS_FF_Z4c}
\end{figure}

As first test, we consider a stationary BS with a free-field potential setting $\mu=1$
and $|\phi_c|=0.01$. This configuration has been studied in the literature already,
e.g.~\cite{Palenzuela:2006wp}.
We present resolution effects, Fig.~\ref{fig:single_BS_FF_res},
the influence of the time integrator, the order of the spatial finite differences,
and the Courant-Friedrichs-Lewy factor,
Fig.~\ref{fig:single_BS_FF_RKcfl}, and of the constraint damping scheme of the Z4c
evolution system, Fig.~\ref{fig:single_BS_FF_Z4c}.

Convergence is studied by evolving the setup with three different resolutions,
see Tab.~\ref{tab:tests-grid}, and we report results in
Fig.~\ref{fig:single_BS_FF_res}. The top panel shows for illustrative purposes the
real part of $\phi$ at the center of the star for BS$_{\rm static}^{\rm FF}$-R3,
the middle panel shows the difference of the amplitude $\phi$ of the center of the star
as a function of time $\Delta |\phi_c|/|\phi_c| = |\phi(r=0,t)|/|\phi(r=0,t=0)|-1$, and the bottom panel
reports the Hamiltonian constraint $\mathcal{H}$.
The simulations are chosen in such a way that resolution is increased by a factor of
$2$ from one setup to the next one.
We find that for $\Delta x = 0.250$ (resolution R1) the central amplitude of the scalar field
is decreasing over time and after $t=2000$ (not shown in the figure)
is $\sim 20\%$ smaller than initially, while resolution $R2$ has only a $5\%$ error.
Overall, the error in the central amplitude of the scalar field
shows clear second order convergence; see dashed lines.
Similarly, also the Hamiltonian constraint (third panel) is
second order convergent. Note further that we do not solve the CTS equations 
to obtain the initial configurations, but only employ the methods discussed above for the 
construction of single BS spacetimes. This is the reason for the initial transient phase 
before $t=100$.

To further assess the robustness of our BS simulations
and to find parameters best suited for efficient evolutions,
we test the influence of different time integrators,
different Courant-Friedrichs-Lewy (CFL) factors, and different orders for the
spatial finite differencing stencils; see
Fig.~\ref{fig:single_BS_FF_RKcfl}.
Regarding the impact of the order of the spatial derivatives,
we find that between the default setup, which employs 4th order
finite differences, and sixth order finite differencing stencils (FD6)
no noticeable difference is present. In contrast, varying the
order of the explicit Runge-Kutta scheme
between 3rd order (RK3) and 4th order (RK4) has a large effect.
Indeed the error in the central density is about a factor of 4 larger and the
Hamiltonian constraint increases by more than an order
or magnitude if RK3 is employed.
Finally, the high dependence on the exact settings of the time integration
is also assessable by varying the CFL condition. Varying the CFL factor between
$c=0.1,0.2,0.3$, and $0.25$ for the default setup, we find that an increase of the
CFL factor by $0.05$ leads to an increase in $\Delta |\phi_c|/|\phi_c|$ by a factor of $2$.
This observation implies that,
since the computational costs depend only linearly on the CFL condition,
BS simulations should employ smaller CFL conditions than default NS simulations.

As a final test we also want to investigate the possible imprint of the
Z4c damping parameter $\kappa_1$ on the simulation of the isolated BS.
As shown in Ref.~\cite{Weyhausen:2011cg}
a damping parameter chosen too large can effect the physical properties.
In Fig.~\ref{fig:single_BS_FF_Z4c} we study the effect of $\kappa_1$ on
the central value of the scalar field amplitude and on the Hamiltonian
constraint violation, as typically for NS simulations we set $\kappa_2=0$.
We find overall that values for $\kappa_1 \in [0,0.1]$ do not
affect $\Delta |\phi_c|/|\phi_c|$. However, surprisingly, the constraint violation
is \textit{not} reduced for an increasing damping parameter.
The smallest constraint violation is obtained for $\kappa_1=0.0$.

\subsection{Stationary boson stars: solitonic potential}

\begin{figure}[t]
\centering
\includegraphics[width=1.\textwidth]{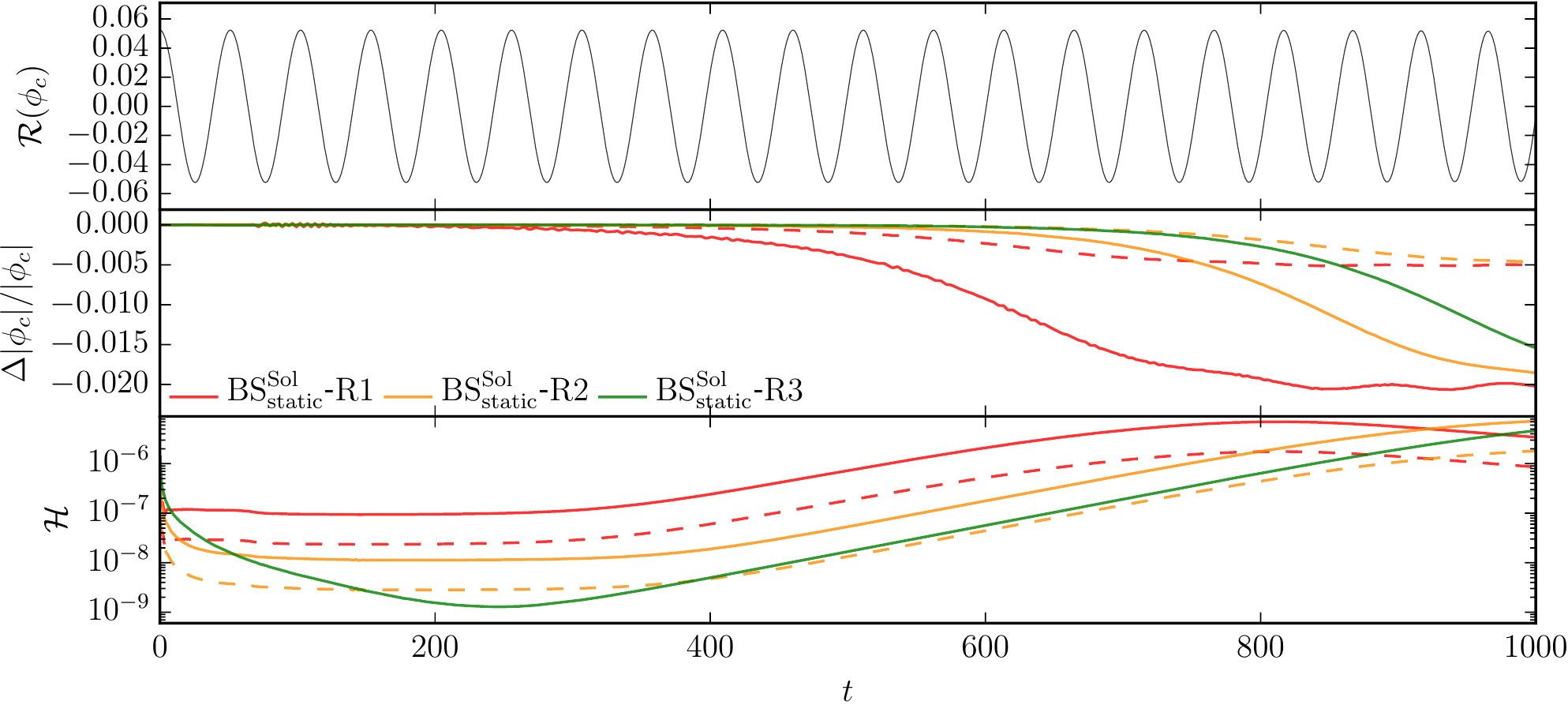}
\caption{Effect of resolution on the isolated solitonic BS.
From top to bottom:
central value of the real part of the scalar field $\phi_c$,
relative difference of the scalar field amplitude at the
center of the star,
Hamiltonian constraint.
Different colors refer to different resolutions,
Tab.~\ref{tab:tests-grid}. Dashed lines show rescaled results assuming second order convergence.}
\label{fig:single_BS_S_res}
\end{figure}

Let us proceed by studying single solitonic BS configurations.
We pick a configuration with a central field value $\phi_c\approx0.0523$
as in~\cite{Macedo:2013jja}.
Figure~\ref{fig:single_BS_S_res} analyses the imprint of resolution.
We clearly find that for low resolutions (R1) the central
value of the scalar field amplitude drops by about $2\%$
during the course of the evolution up to $t=1000$.
A similar behavior is true for R2 and R3.
Overall, the steep gradients near the ``surface''
of the BS, Fig.~\ref{fig:profile_single_stars}, get smeared out
and the system relaxes to a stable configuration.
The transition takes longer for higher resolutions,
but is present for all studied cases.
Similarly the Hamiltonian constraint increases until the new configuration
with less steep gradients is obtained.
Considering resolutions R2 and R3, we find approximately
second order convergence in the late time of the simulation,
but in general the convergence properties are worse than for the free-field BSs.

We do find that, although generally larger,
the Hamiltonian constraint for the studied solitonic BS
does not show a noticeable dependence on the CFL condition
as the free-field potential.
We suggest two reasons for this observation
(i) the free-field BS has a oscillation frequency significantly higher that the studied
solitonic star case;
(ii) the leading error source for the evolution of the solitonic stars
is connected to the large gradient in $\phi$, and $E^{(\phi)}$ near the star's
``surface''.

These suggestions are supported by the observation that the
order of the time integrator (RK3 and RK4) also does not effect the evolution.
In contrast, the 6th order finite differencing stencil
does lead to a smaller Hamiltonian constraint violation and
a smaller difference in the central amplitude of the scalar field.
Unfortunately, using higher order finite difference stencils
not only increases the necessary computational costs
to evaluate the right hand side of the evolution equations,
but also necessitates larger mpi-buffer and mesh refinement-buffer
zones~\cite{Dietrich:2015iva} which lowers the scalability of the
numerical code and results in larger memory requirements and communication
for the simulations.

\begin{figure}[t]
\centering
\includegraphics[width=1.\textwidth]{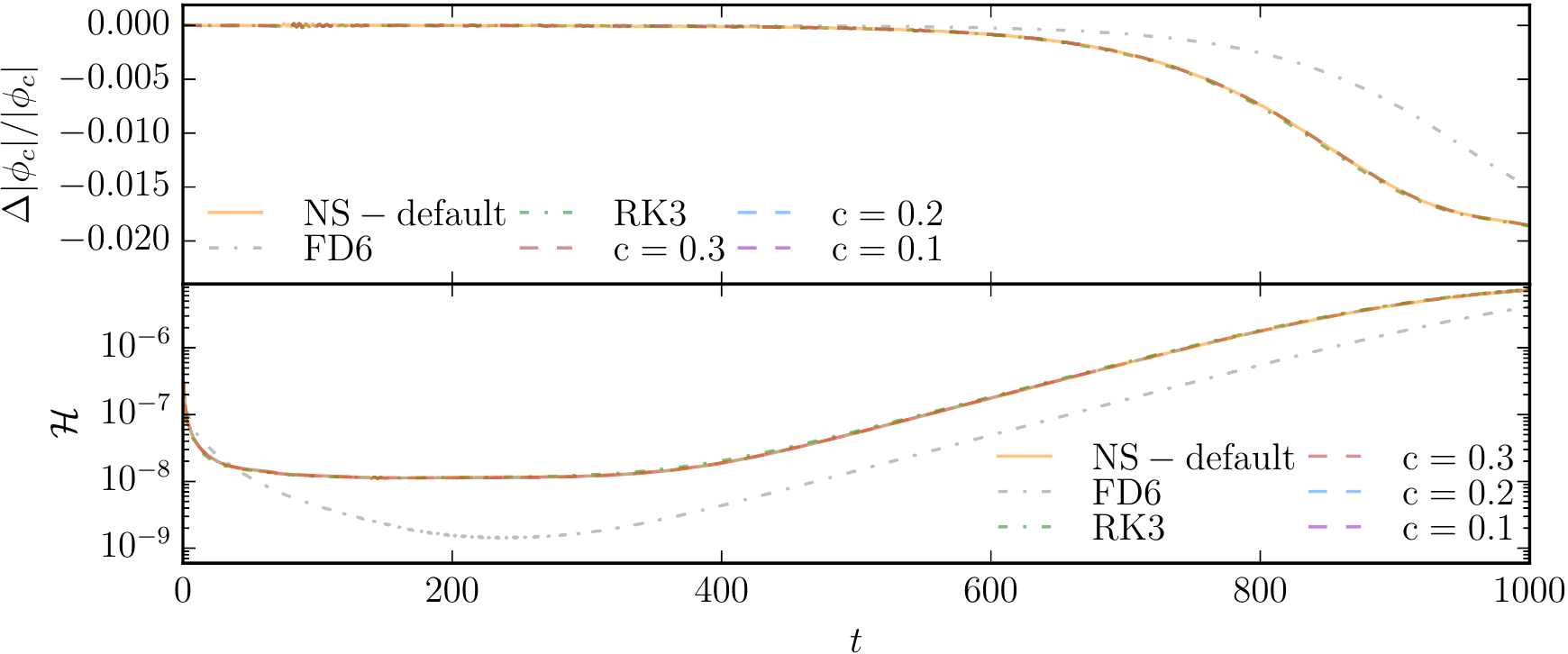}
\caption{Effect of the time integrator, CFL condition, and order of the spatial derivatives
on the central amplitude of the scalar field (top) and Hamiltonian constraint (bottom) for the solitonic 
BS case.
As a default setup, we use a 4th order Runge-Kutta scheme with a CFL condition
of $c=0.25$ with 4th order finite differencing stencils for the spatial derivatives,
we refer to this as 'default', since it refers to our default choice for the simulation
of NS spacetimes. We employ the $\rm R2$ resolution.}
\label{fig:single_BS_S_RKcfl}
\end{figure}

\begin{figure}[t]
\centering
\includegraphics[width=1.\textwidth]{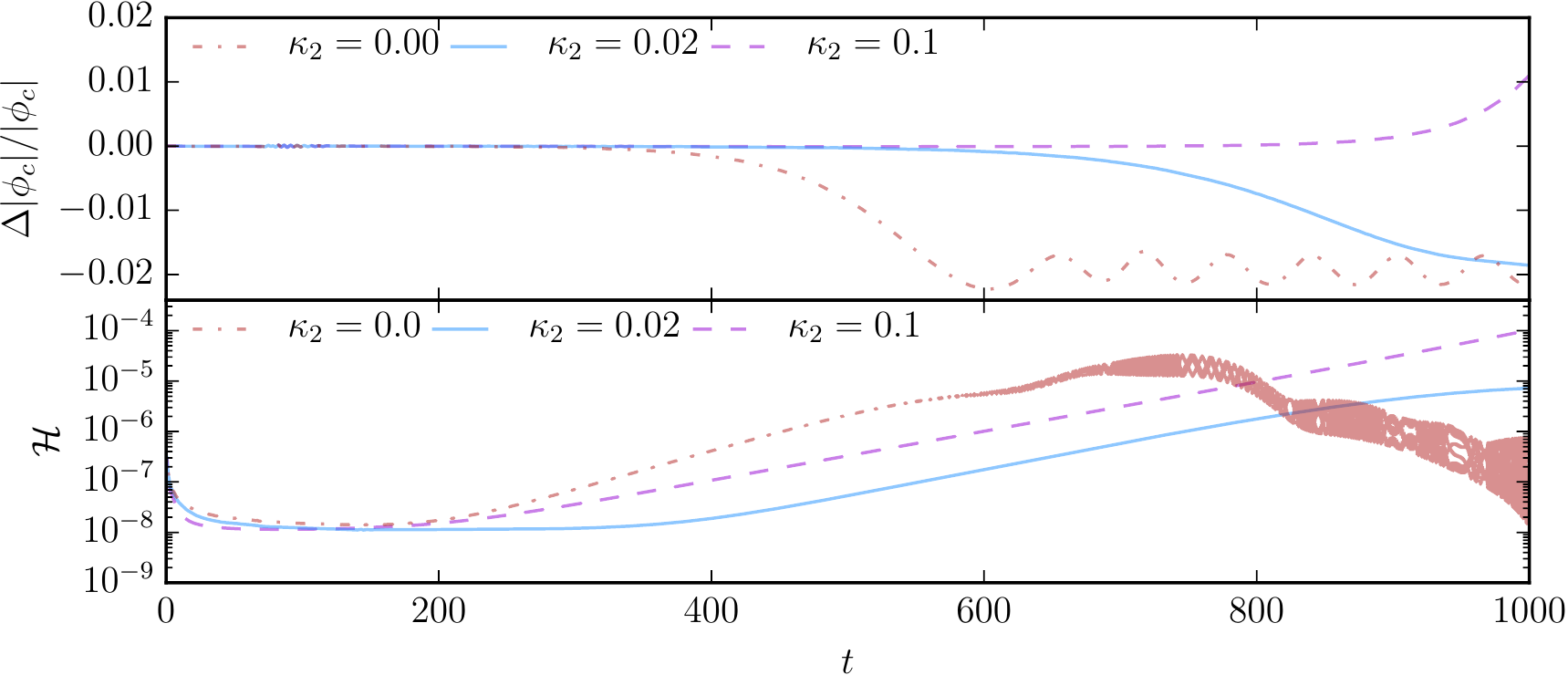}
\caption{Influence of the $\kappa_1$ damping parameter on the central value of $|\phi|$ (top panel)
and on the Hamiltonian constraint (bottom panel). We set for all tests $\kappa_1=0$.}
\label{fig:single_BS_S_Z4c}
\end{figure}

Finally, we want to address the impact of the Z4c-damping parameter
on the solitonic BS.
While we found for the free-field BS that the larger the damping parameter,
the smaller the Hamiltonian constraint and that
$\Delta |\phi_c|/|\phi_c|$
was basically independent of $\kappa_1$, neither observation is
true for the studied solitonic stars.
Indeed we find that for $\kappa_1=0.02$ the Hamiltonian
constraint is smallest and also that for this damping parameter
$\Delta |\phi_c|/|\phi_c|$ is smallest.
Interestingly, while for $\kappa_1=0.0$ the evolution
is stable, and settles to a state with central scalar field values about $\sim 2\%$
smaller than the initial configuration, cf.~discussion above.
The simulation with $\kappa_1=0.1$ becomes unstable
and the central density increases significantly and leads
to a failure of the simulation at later times ($t>1000$)
not shown in Fig.~\ref{fig:single_BS_S_Z4c}.

\subsection{Stationary neutron stars}

\begin{figure}[t]
\centering
\includegraphics[width=0.95\textwidth]{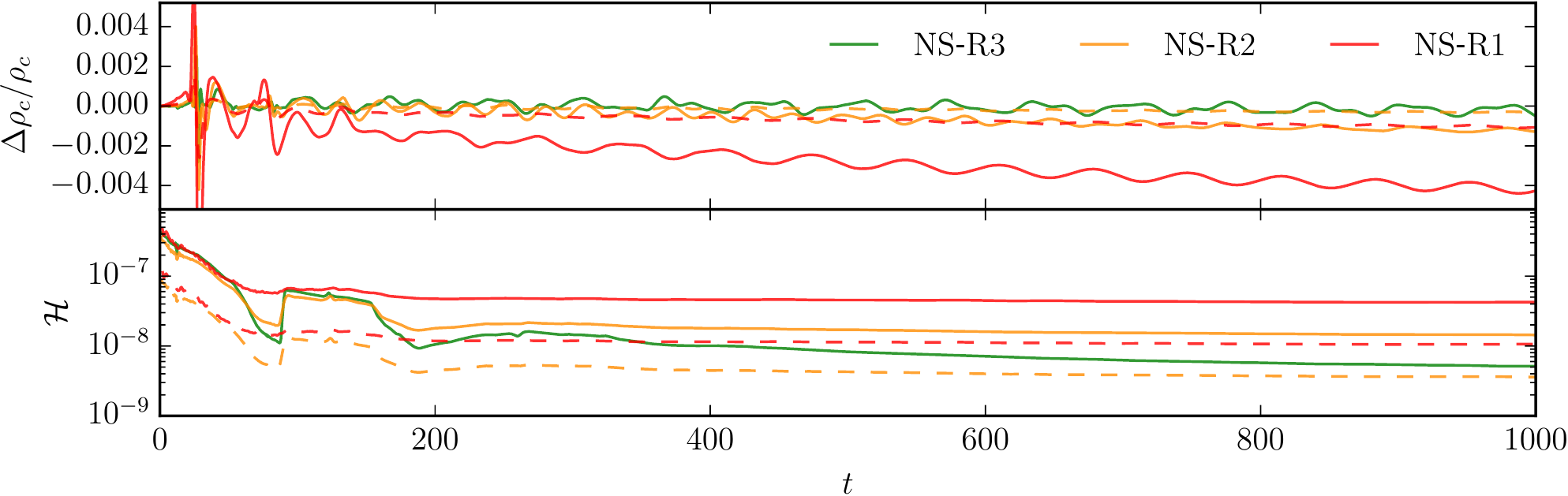}
\caption{Effect of resolution for isolated NS simulations.
Top panel:
difference in the central density $\rho$ over time;
Bottom Panel: Hamiltonian constraint violation over time.
Different colors refer to different resolutions,
Tab.~\ref{tab:tests-grid} and dashed lines
are rescaled assuming second order convergence.}
\label{fig:single_NS_res}
\end{figure}

Although already studied in detail in Refs.~\cite{Thierfelder:2011yi,Dietrich:2015iva,Bernuzzi:2016pie},
we present some simulations of single, static NSs for comparison
with the previously presented BS configurations.
This will allow us to find optimal parameters for the dynamical
simulation of mixed BSNS systems.

Figure~\ref{fig:single_NS_res} shows the central density
oscillation and the Hamiltonian constraint for the resolutions R1,R2,R3; Tab.~\ref{tab:tests-grid}.
The error in the central density $\Delta \rho_c/\rho_c$
is significantly smaller than for the BS simulations.
We assume that the main reason for the good conservation of
$\rho_c$ is that in contrast to the BS case the NS's matter fields
are in principle stationary although small numerical uncertainties
lead to a non-trivial evolution.
Considering the dashed lines in the top panel of Fig.~\ref{fig:single_NS_res},
we find that the central density converges with second order as already pointed
out in Refs.~\cite{Thierfelder:2011yi,Bernuzzi:2016pie}.
Additionally, also the Hamiltonian constraint decreases over time,
but does not show a clean convergence order,
which again is in agreement with previous simulations
for the chosen LLF flux method~\cite{Bernuzzi:2016pie}.

\begin{figure}[t]
\centering
\includegraphics[width=0.95\textwidth]{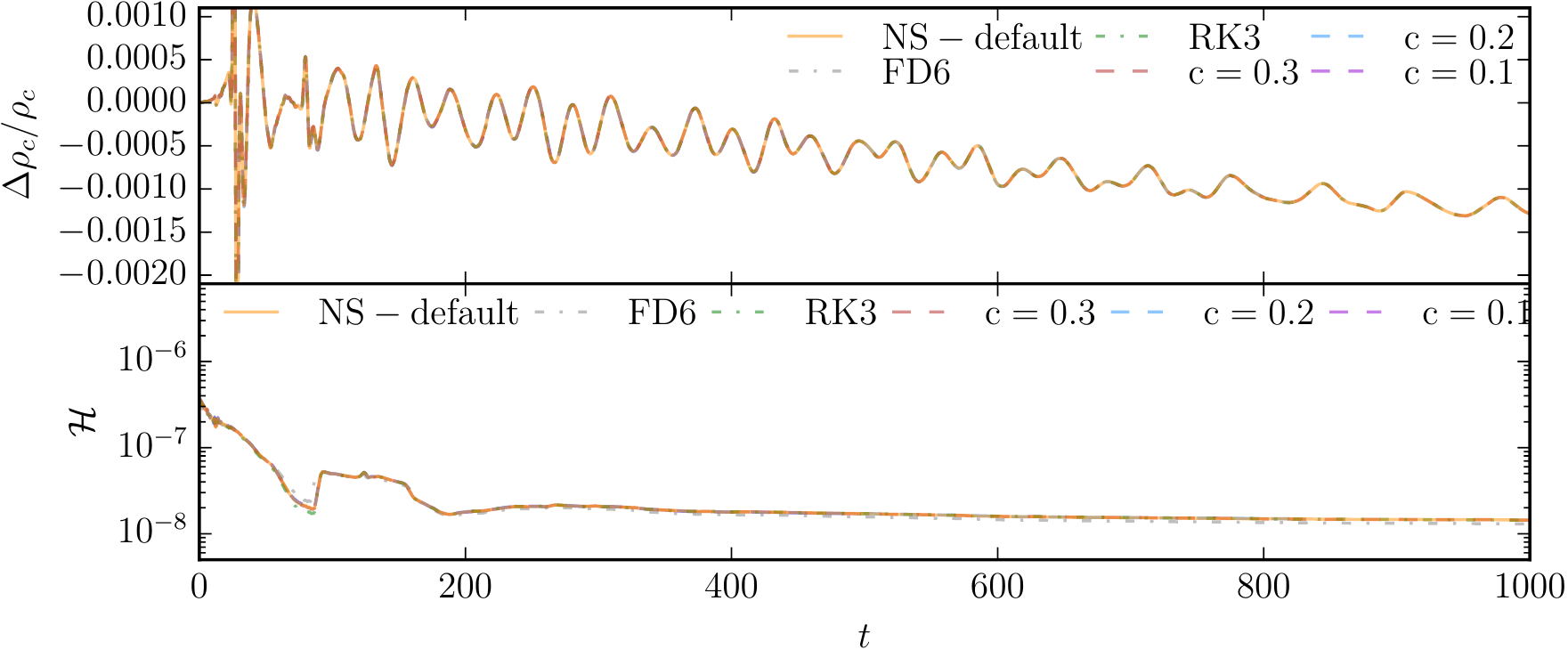}
\caption{Effect of the time integrator,
CFL condition, and order of the spatial derivatives of the spacetime variables.
We show the difference in the central density (top panel)
and the Hamiltonian constraint (bottom panel).
As a default setup, we use a 4th order Runge-Kutta scheme
with a CFL condition of $c=0.25$ and employ 4th order
finite differencing stencils for the spacetime variables;
we refer to this setup as 'default'.
We employ the $\rm R2$ resolution.}
\label{fig:single_NS_RKcfl}
\end{figure}

Considering the impact of the CFL condition,
the order of the finite differencing scheme,
and the employed order of the explicit Runge-Kutta scheme;
we find that $\Delta \rho_c/\rho_c$ and $\mathcal{H}$
are unaffected by the exact settings (Fig.~\ref{fig:single_NS_RKcfl}).
However, we stress that NS simulations
depend crucially on the numerical flux, the flux limiter, as
well as the employed atmosphere setup. Since the particular
choice of the numerical flux etc.\ does not
effect the evolution of the bosonic fields, we will simply refer
here to the studies of~\cite{Bernuzzi:2012ci,Radice:2013hxh,
Radice:2013xpa,Guercilena:2016fdl,Bernuzzi:2016pie}.

Finally, considering the influence of the Z4c
damping parameter $\kappa_1$,
we find that the smallest constraint violation is
obtained for $\kappa_1=0.00$ and $\kappa_1=0.02$,
see also~\cite{Weyhausen:2011cg} for more details.

\begin{figure}[t]
\centering
\includegraphics[width=0.95\textwidth]{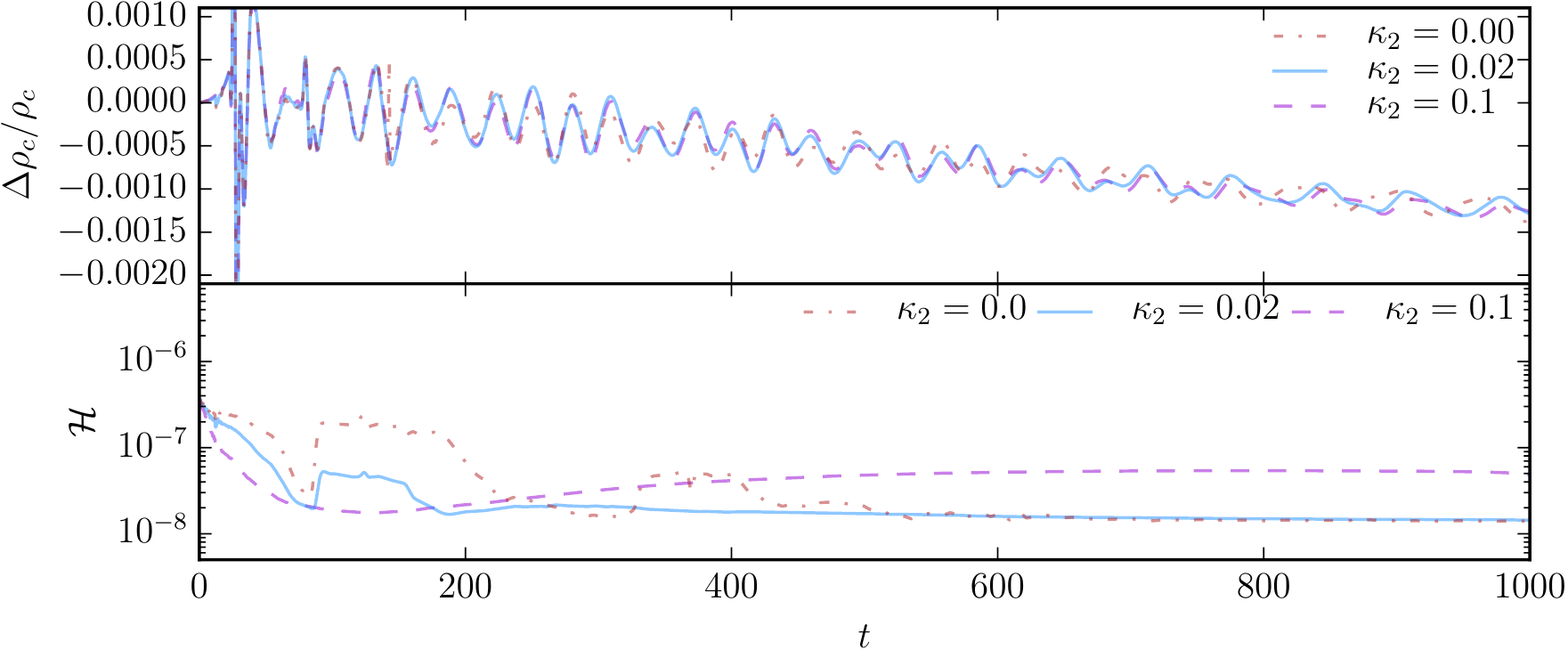}
\caption{Influence of the $\kappa_1$ damping parameter on the
central density of the NS (top panel)
and on the Hamiltonian constraint (bottom panel).
We set for all tests $\kappa_1=0$ and employ $\rm R2$ resolution.}
\label{fig:single_NS_Z4}
\end{figure}

\subsection{Boosted boson stars}

\begin{figure}[t]
\centering
\includegraphics[width=0.95\textwidth]{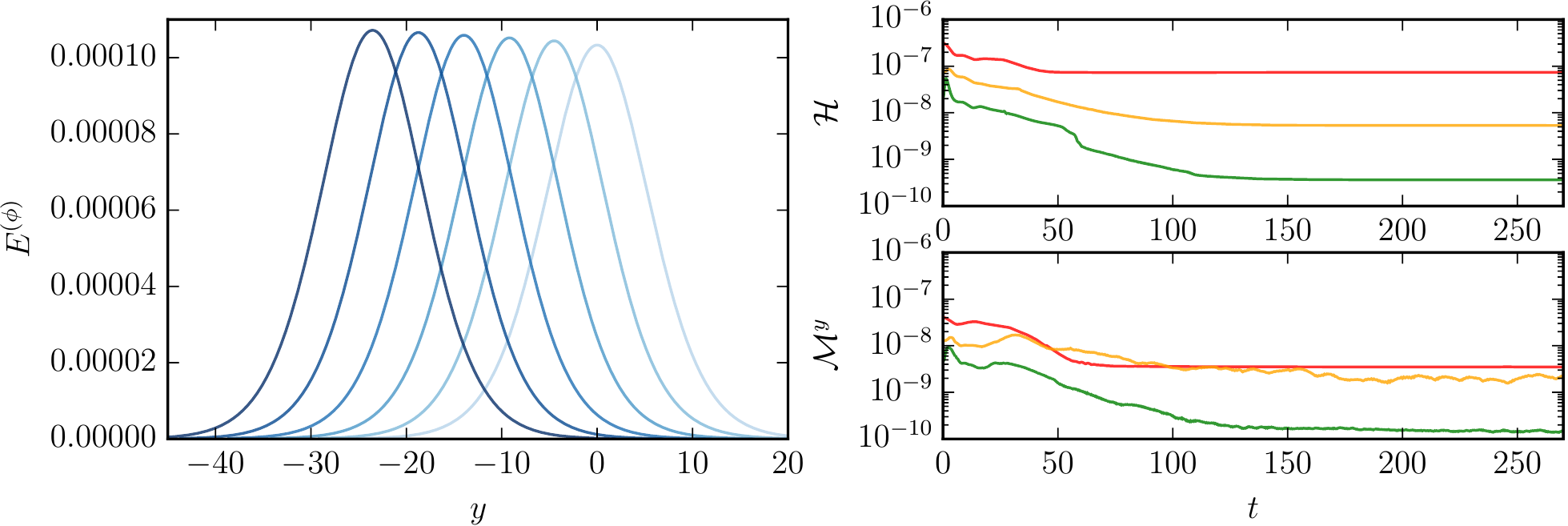}
\caption{Left panel: Profile of the energy density of the boosted BS configuration
{BS$_{\rm boost}^{\rm FF}$-R3}. We show different instances of time corresponding to
$t=0,50,100,150,200,250$ (order by decreasing brightness).
Right panel: Hamiltonian constraint for different resolutions of the {BS$_{\rm boost}^{\rm FF}$}
setup.}
\label{fig:boost}
\end{figure}

In addition to static BSs, we also perform tests for boosted BSs
which will be of particular importance for setting up orbiting BSNS systems.
After boosting the scalar field variables we obtain the spacetime variables
by solving the CTS equations.

We present consistency checks for the boosted BSs in Fig.~\ref{fig:boost},
where we show the energy density profile in the left panel for different time slices
corresponding to $t=0,50,100,150,200,250$.
The right panels of Fig.~\ref{fig:boost}
show the Hamiltonian constraint (top) and the y-component of the momentum
constraint (bottom).
It is clearly visible that with increasing resolution the constraints decrease
and that no artificial deformation of the energy density profile occurs
during the time evolution.

\subsection{Single Star Summary}

Finally, we wish to summarize our single star observations to
motivate our choice for the ensuing binary configurations.
As seen from the free-field BSs, the exact choice of the CFL condition
has a large influence on the dynamical evolution.
Since the computational costs depend only linearly
on the CFL condition, we decide for the following to lower the CFL
condition to $c=0.15$, with respect to the NS-default of $c=0.25$.
Despite this change, we use our NS-default
setup of RK4, 4th order finite difference stencils for the metric and bosonic field
variables, and $\kappa_1=0.02$.
Those settings have been shown to permit
robust BS and NS simulations.

\section{Testbeds: Head-on collision of binary boson star systems}
\label{sec:tests:binary}

As a next test, we will simulate BS-BS systems
and compute the GW signal emitted during their head-on collision.
Both stars are separated at a distance of $d=60$. The individual configurations
are equivalent to ${\rm BS}_{\rm static}^{\rm FF}$, i.e.,
the individual masses are $M_A=M_B=0.36$.
The stars have no initial velocity.

\begin{figure}[t]
\centering
\includegraphics[width=0.95\textwidth]{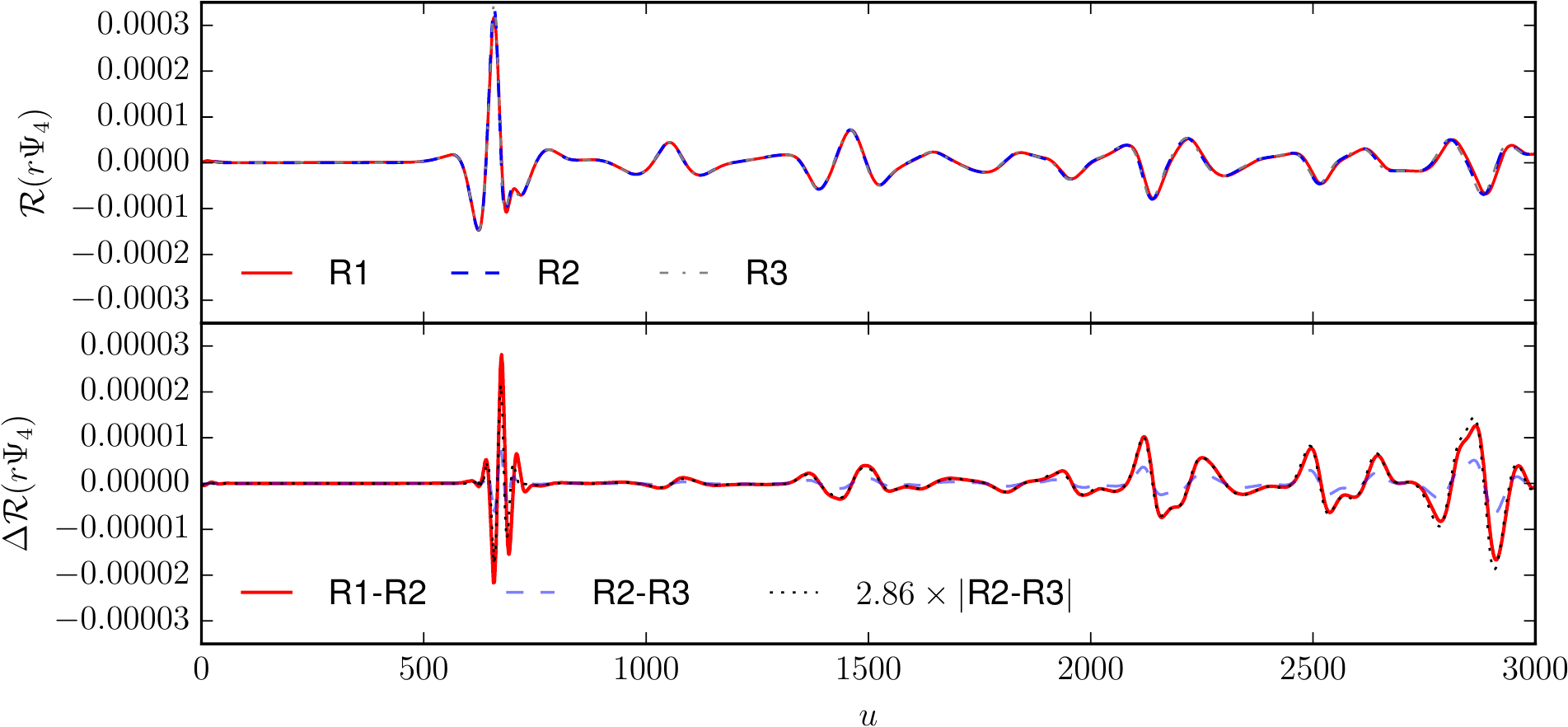}
\caption{Real part of the (2,2)-mode of the curvature multipole for different resolutions labeled by R1, R2, R3
with increasing resolution; see text for details. In the bottom panel we present the differences between
different resolutions and rescale the difference between R2-R3 according to second order convergence (black line).}
\label{fig:BSBS_wavesBAM}
\end{figure}

For the \BAM evolution we employ three different resolutions
consisting of $7$ refinement levels and a finest grid spacing of $0.125, 0.083, 0.0625$.
The number of points in the moving mesh refinement boxes is equal to
$128,192,256$ and in the non-moving levels $160,240,320$, respectively.
We present the (2,2)-mode of the curvature multipole $\Psi_4$
in Fig.~\ref{fig:BSBS_wavesBAM} for the three different resolutions
as a function of the retarded time
\begin{equation}
u = t-r_{\rm ext}-2M\log\left(\frac{r_{\rm ext}}{2M}-1\right),
\end{equation}
with the extraction radius $r_{\rm ext}$ and the total mass of the system $M=M^A+M^B$.

The GW signal (top panel of Fig.~\ref{fig:BSBS_wavesBAM}) consists of two phases:
(i) The merger signal released during the head-on collision of the two stars.
The signal's qualitative shape is similar to the head-on collision of other
compact objects as BHs or NSs~\cite{Anninos:1993zj,Anninos:1994gp}.
(ii) The postmerger signal emitted from the merger remnant.
Considering the evolution of the different resolution setups, we find that indeed
all resolutions seem to resolve properly the GW signals, while
a dephasing is visible over longer times at the end of the simulation.
The dephasing and the overall error (bottom panel) shows a clear 2nd order convergence.
To aid visualization, we rescale the phase difference between R2 and R3 (black line), which agrees
after rescaling with the difference between the setups R1 and R2 (red line).
We should emphasize that the clean 2nd order convergence after the merger is something which is
not achieved in any BNS simulation due to the formation of shocks. In contrast, BSs 
do not create shocks and discontinuities in the field variables which enables clean 
convergence beyond the merger.

\begin{figure}[t]
\centering
\includegraphics[width=0.95\textwidth]{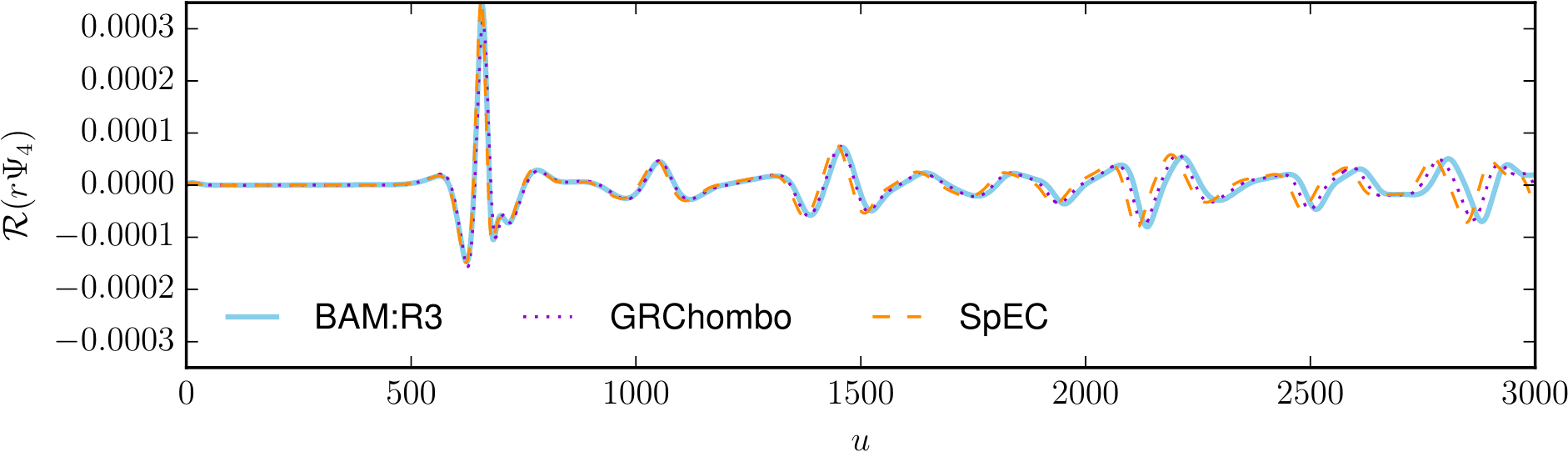}
\caption{
Real part of the (2,2)-mode of the curvature multipole for different numerical codes.
We compare the \BAM code (blue) with the GRChombo code~\cite{Clough:2015sqa} (violet) and the SpEC code~\cite{SpEC} (orange).
Overall good agreement between all three codes is obtained.}
\label{fig:BSBS_code}
\end{figure}

In addition to the resolution study presented above, we also
compare the \BAM results with evolutions obtained with
SpEC~\cite{SpEC} and GRChombo~\cite{Clough:2015sqa}.
Let us point out that although SpEC and GRChombo allow for the evolution of BSs,
they are currently unable to evolve BS and NS mergers in mixed systems.

GRChombo applies similar methods as the BAM code, with the main differences being:
the use of the CCZ4 formalism~\cite{Alic:2011gg,Bona:2003fj}
and a fully adaptive mesh scheme with non-trivial nesting
topologies according to the Berger-Rigoutsos
block-structured adaptive mesh
algorithm~\cite{BergerRigoutsis91}.
In the BS-BS evolution we allow up to
7 levels of 2:1 refinement above a base grid of
length $1024$ with $128^3$ points, and a
CFL factor of $0.1$.
The dynamical regridding is based on local derivatives
in the bosonic fields - not all levels are utilized throughout -
and the Weyl scalar is extracted on a sphere at a radius of $r_{\rm ext}=250$.

SpEC is a multi-domain pseudo-spectral collocation code that uses the method of lines to evolve
a first order formulation~\cite{Lindblom:2005qh} of the Einstein field equations in the generalized
harmonic gauge~\cite{Pretorius:2005gq, Friedrich1985}. Numerically stable evolutions are achieved by using a
damped harmonic gauge as described in detail in Ref~\cite{Lindblom:2009tu}. Adaptive mesh refinement is implemented by simultaneously adjusting the number of subdomains as well as the order of the spectral basis functions
in each subdomain see~\cite{Szilagyi:2014fna}.  For the simulations shown here, the target
truncation error on metric variables is set to be $10^{-4}/e$. The gravitational wave is extracted at a coordinate
radius of $r_{\rm ext}=540$.  

Figure~\ref{fig:BSBS_code} shows the time evolution of the GW signal
for the three different codes. Overall we find very good agreement. 
The small differences might be caused by slightly different 
initial configurations, since all three codes solve the Einstein Constraints 
with different techniques. 
Further studies are planned to allow for a better error budget of BS simulations 
including extensive tests and analyses for GRChombo and SpEC. 

The overall agreement between the three codes is of particular importance
not only since it validates the implementation
of the \BAM routines, it is also (to the best of our knowledge) the first direct comparison
between different NR codes performing simulations of ECOs.
This shows that NR has arrived at a stage where reliable
simulations of ECOs are within reach.
Although the current test is limited by the fact
that the stars perform head-on collisions, it is a starting point for further
usage of these three code for the study of ECOs in the context of GW modeling.

\section{Mixed Binaries: boson star -- neutron star mergers}
\label{sec:BSNS}

\begin{table}[t]
  \centering
  \caption{Overview of mixed binary evolutions.
    The columns refer to:
    the simulation name,
    the central field value of BS,
    the mass of the gravitational BS (in isolation), 
    the mass of the gravitational NS (in isolation), 
    the number of points in the fixed (moving) boxes $n$ $(n^{mv})$,
    the grid spacing in level $l=0,L-1$ ($h_{0},h_{L-1}$).
    The resolution in level $l$ is $h_l=h_0/2^l$. For all simulations
    we employ a total of seven refinement levels $(L=6)$.
    The three outermost levels are fixed employing $n^3$ grid points,
    the remaining levels employ $(n^{mv})^3$ points. All simulations
    make use of bitant symmetry to reduce computational costs.
    Note that setup BSNS$_{0.01}^{\rm head}$-R4 is only employed to test 
    the multigrid solver and not evolved. }
  \begin{tabular}{l|ccc|cccc}
     name              & $\phi_c$ & $M_{\rm BS}$ & $M_{\rm NS}$  & $n$ & $n^{mv}$ & $h_0$ & $h_{L-1}$  \\
     \hline
     BSNS$_{0.005}^{\rm head}$-R3 & $0.005$& $0.264$ & $1.350$ & $192$ & $360$ & $8.0$  & $0.125$ \\
     \hline
     BSNS$_{0.01}^{\rm head}$-R1  & $0.01$ & $0.361$ & $1.350$  & $96$  & $180$  & $16.0$ & $0.250$\\
     BSNS$_{0.01}^{\rm head}$-R2  & $0.01$ & $0.361$ & $1.350$  & $128$ & $240$ & $12.0$ & $0.1875$\\
     BSNS$_{0.01}^{\rm head}$-R3  & $0.01$ & $0.361$ & $1.350$  & $192$ & $360$ & $8.0$  & $0.125$ \\
     BSNS$_{0.01}^{\rm head}$-R4  & $0.01$ & $0.361$ & $1.350$  & $256$ & $480$ & $6.0$  & $0.09375$ \\
     \hline
     BSNS$_{0.02}^{\rm head}$-R3  & $0.02$ & $0.475$ & $1.350$ & $192$ & $360$ & $8.0$  & $0.125$\\
     \hline \hline
     BSNS$_{0.08}^{\rm orb}$-R0   & $0.08$ & $0.633$ & $1.350$ & $120$ & $64$ & $24.0$  & $0.375$  \\
  \end{tabular}
 \label{tab:BSNS:cases}
\end{table}

\subsection{Head-on collision of boson star -- neutron star binaries}

\subsubsection{Initial data}

\begin{figure}[t]
\centering
\includegraphics[width=0.95\textwidth]{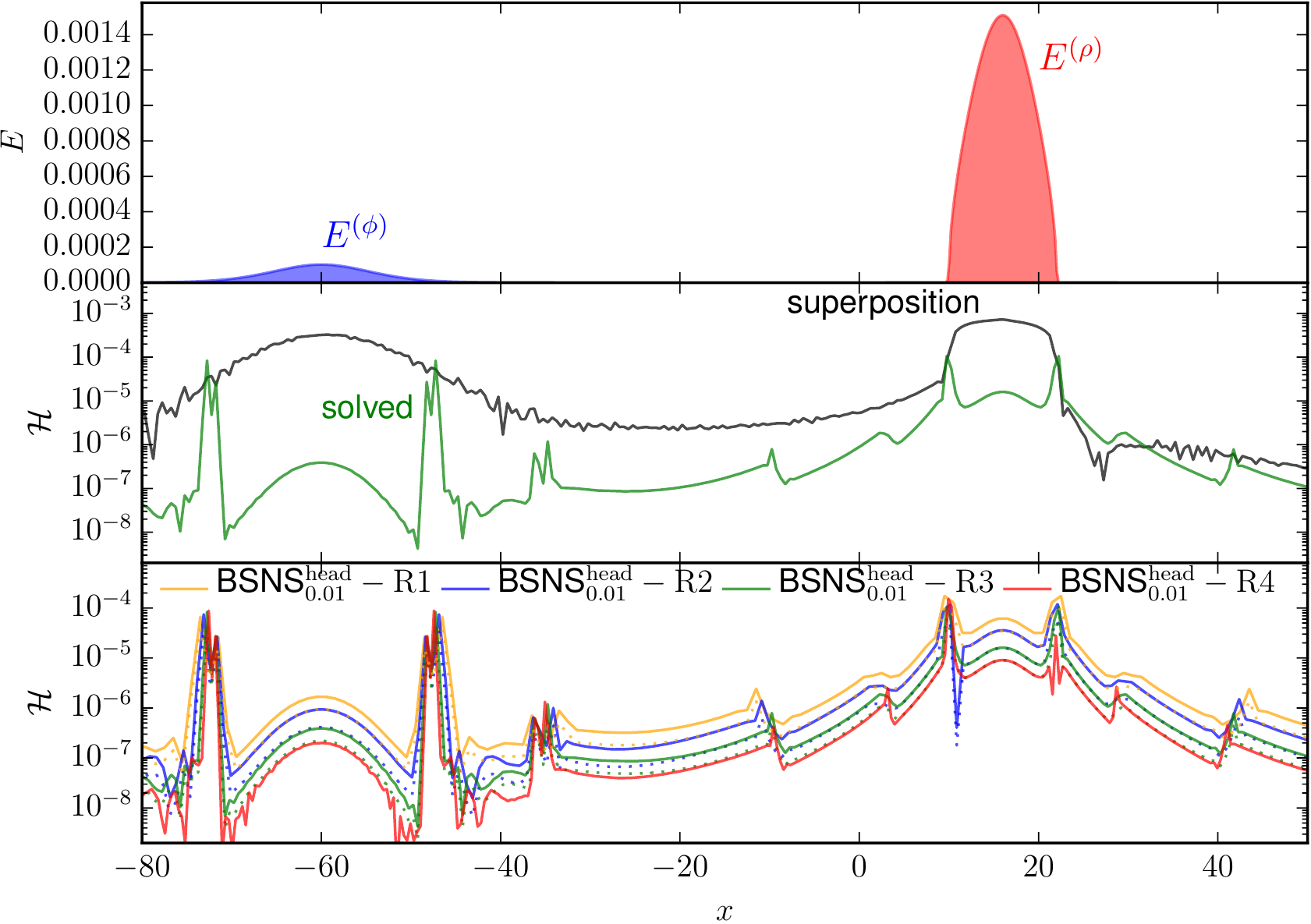}
\caption{Head-on BSNS configuration with component masses of $1.35$.
The top panel shows the energy density $E$ for the bosonic matter (blue) and the baryonic matter (red).
The bottom panel shows the Hamiltonian constraint for the setup after simple superposition (black)
and after resolving with \BAM's multigrid solver.}
\label{fig:multigrid_BSNS}
\end{figure}

As a starting point for the simulation of mixed systems,
we present a BS - NS head-on collision for three different physical cases.
Simulation details can be found in Tab.~\ref{tab:BSNS:cases}.
We place the NS initially at $x=16$ and the BS at $x=-60$.
This way the center of mass for case BSNS$_{0.01}^{\rm head}$ is at the origin
of the numerical domain.
We construct the initial data following the procedure outlined in Sec.~\ref{sec:equations:ID}
using \BAM's multigrid solver.
We show for BSNS$_{0.01}^{\rm head}$-R3
the energy density (blue for bosonic matter and red for baryonic matter)
in the top panel of Fig.~\ref{fig:multigrid_BSNS}.
The middle panel shows the Hamiltonian constraint
for the superposed setup (black) and the setup after solving the CTS equations (green)
employing resolution R3; see Tab.~\ref{tab:BSNS:cases}.
With respect to the superposed configuration the Hamiltonian constraint is decreased by 
several orders of magnitude. However, there are noticeable
constraint violations at the refinement boundaries and
also at the stellar surface of the NS.
In the bottom panel of Fig.~\ref{fig:multigrid_BSNS} we present the Hamiltonian constraint
after solving the CTS equations but employing four different resolutions.
We find that overall second order convergence is obtained. For visibility we rescale the 
constraint violations assuming second order convergence and show these as dotted lines
in the panel.

\subsubsection{Evolution}

\begin{figure}[t]
\centering
\includegraphics[width=0.95\textwidth]{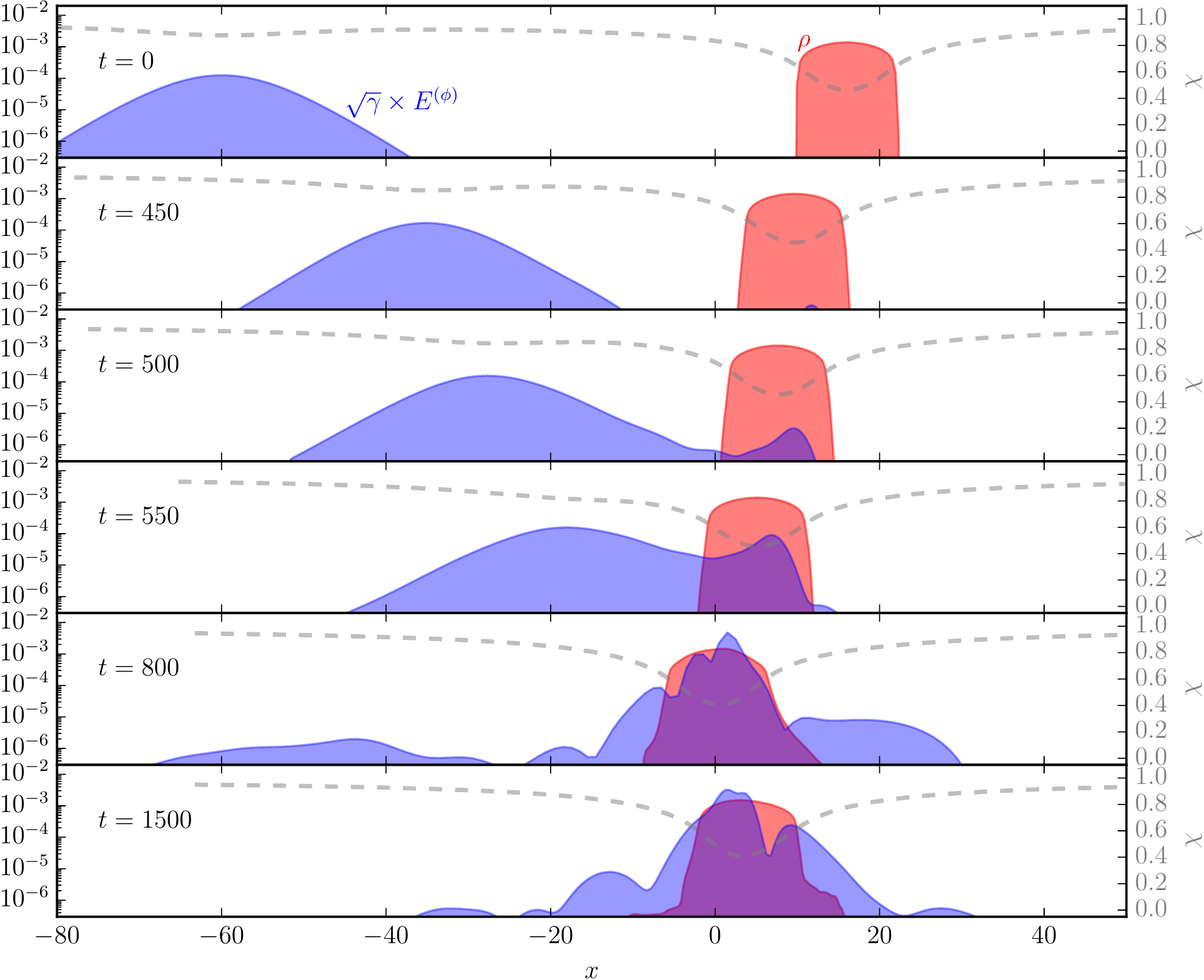}
\caption{Bosonic energy density (blue) and rest mass density of the NS (red) for different
instances of time. Additionally we show the conformal factor as a gray dashed line with
the corresponding axis on the right side. The conformal factor is evaluated on level $l=4$,
i.e., the second to highest resolved level, except for $t=0$ 
where we evaluate it on level $l=3$.}
\label{fig:BSNS_headon}
\end{figure}

\begin{figure}[t]
\centering
\includegraphics[width=0.95\textwidth]{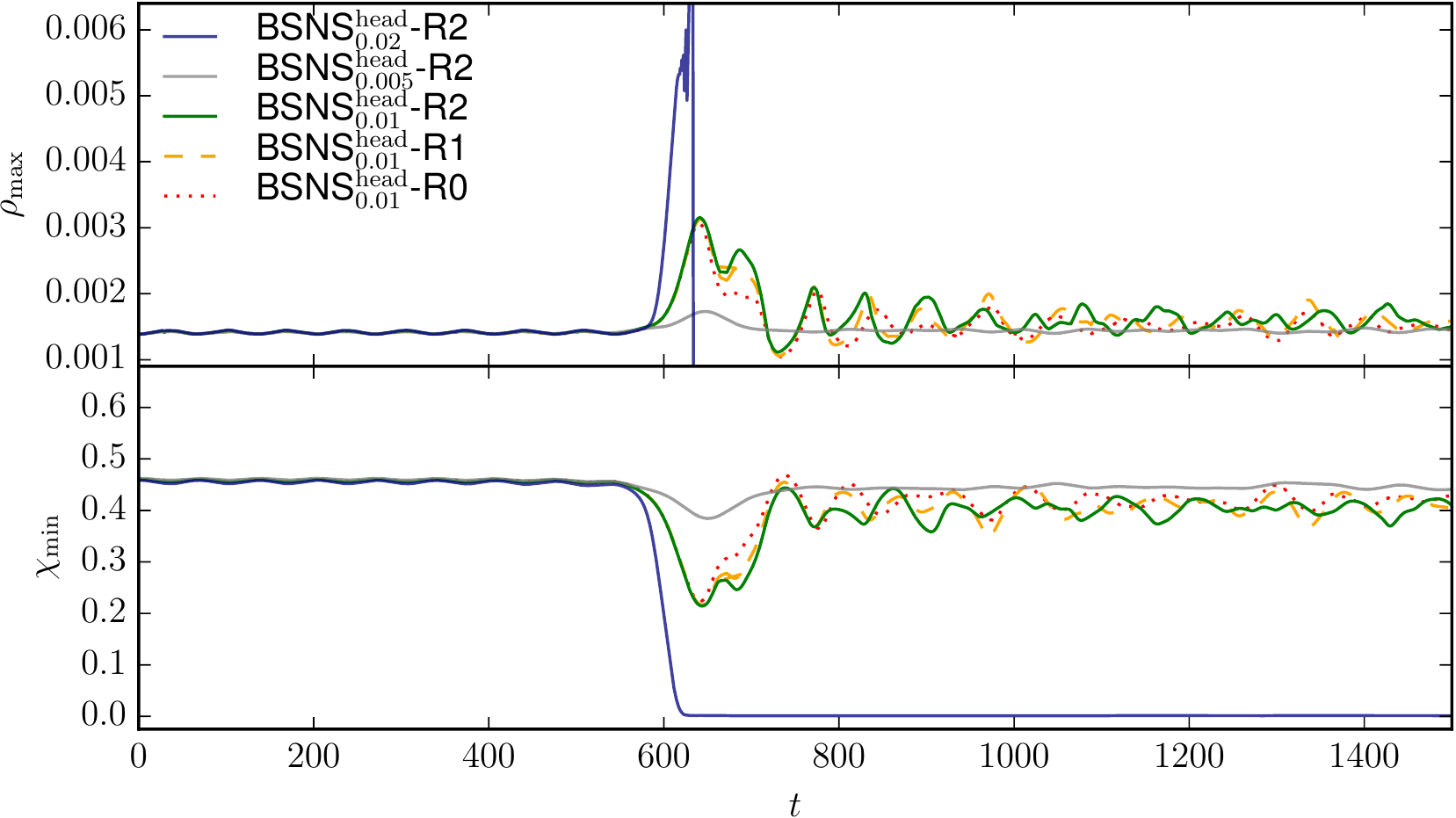}
\caption{Top panel: Maximum rest mass density $\rho_{\rm max}$ of the NS for the cases of Table~\ref{tab:BSNS:cases}. 
Bottom: minimum of the conformal factor $\chi_{\rm min}$ for the NS for the cases of Table~\ref{tab:BSNS:cases}}
\label{fig:BSNS_headon_grhdchi}
\end{figure}

We discuss as an example BSNS$_{0.01}^{\rm head}$-R3 in 
Fig.~\ref{fig:BSNS_headon}. The figure shows the time evolution of the
energy density for the BS~\footnote{We multiply the energy density by $\sqrt{\gamma}$
due to technical reasons to allow for the computation of a ``conserved'' energy density
for the BS similar to the baryonic quantity $D$. Since \BAM also supports axion star 
evolutions there is in general no Noether current which would permit the computation of 
the bosonic mass as a direct equivalent of the baryonic mass for NSs.} (blue)
and the density of the NS (red). Additionally, we show the evolution of the conformal
factor as a gray dashed line; cf.~left axis.
The first panel nicely illustrates that the energy density of the NS 
has a clear surface while the energy density of the BS decays exponentially.
At a time of $t=450$ the initial separation of $76$ has decreased to $45$ 
(where we measure the distance between the minima of the conformal factor $\chi$ inside the 
individual stars). 
At this time a small amount of the bosonic
matter is already present within the NS. This material accumulates in 
the potential well of the NS, but its energy density is still orders of magnitude
below the energy density inside the center of the BS.
At $t=500$ the bosonic energy density increases by about an order of magnitude inside
the NS, the separation between the stars has decreased to $34$ and a clear deformation of
the BS's shape is visible. At about $t=550$ the BS and NS are almost ``merged''. Although hardly
visible on the logarithmic plot, the central density of the NS also starts to increase at this time; 
see top panel of Fig.~\ref{fig:BSNS_headon_grhdchi}.
In contrast, the conformal factor decreases and reaches
a minimum of $\chi_{\rm min} \sim 0.25$ at about $t\approx650$ (see bottom panel of Fig.~\ref{fig:BSNS_headon_grhdchi}),  
but gravitational forces are not strong enough in this case to trigger BH formation. 
At $t=800$ the NS is still deformed and surrounded by a highly dynamical bosonic cloud. 
At $t=1500$ we find that the main bosonic matter surrounds the NS where gravitational 
forces are strongest. The central density reaches values about 
$10$-$20\%$ larger than initially (a similar statement holds for the conformal factor). 
Although the system is close to BH formation no BH forms by the end of the simulation. 
However, increasing the mass of the BS does lead to BH formation. We tested this with 
the evolution of BSNS$_{0.02}^{\rm head}$-R3 (dark blue lines in Fig.~\ref{fig:BSNS_headon_grhdchi}).
On the other hand, if the BS mass is decreased (BSNS$_{0.005}^{\rm head}$-R3) no BH forms and the NS is surrounded 
by a bosonic cloud. Overall for these setups the system stabilized faster than for systems with 
larger BS masses.

\begin{figure}[t]
\centering
\includegraphics[width=0.95\textwidth]{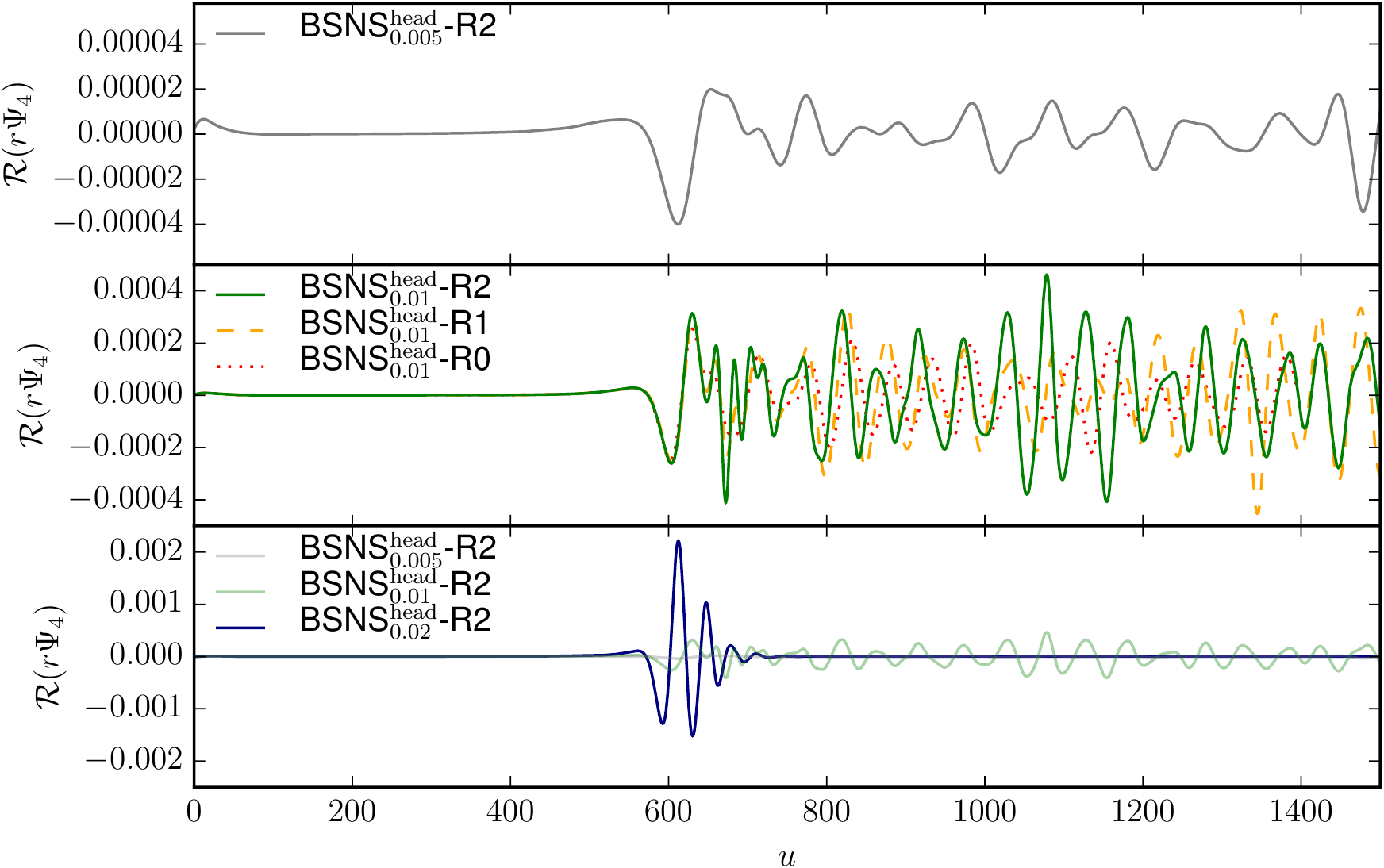}
\caption{(2,2)-mode of the GW signal $\Psi^4$ for the three different head-on configurations of Table~\ref{tab:BSNS:cases}:
BSNS$_{0.005}^{\rm head}$ (top), BSNS$_{0.01}^{\rm head}$ (middle), BSNS$_{0.02}^{\rm head}$ (bottom).
In the middle panel we show simulations for different resolutions to allow 
a qualitative interpretation.
In the bottom panel we also include other configurations (as shaded lines) to allow 
a direct comparison of the amplitude of the emitted GWs.}

\label{fig:BSNS_headon_waves}
\end{figure}

Finally, we present the emitted GW signals for the simulated setups in 
Fig.~\ref{fig:BSNS_headon_waves}. The different panels show the three different physical scenarios with 
different central maximum values of the scalar field. 
In the middle panel we include all three simulated resolutions 
for setup BSNS$_{0.01}^{\rm head}$. While initially the signal shows a similar quantitative behaviour, 
we find that at least after $u=1000$ resolutions R1 and R2 show clear differences. 
This is in contrast to the simulations of the BBS setups, but is to be expected due 
to discontinuities for GRHD simulations. 
Considering the difference between different physical setups, we find that the 
BSNS$_{0.005}^{\rm head}$ has a very small GW amplitude compared to the two other cases 
(see bottom panel of Fig.~\ref{fig:BSNS_headon_waves}). 

\subsection{Boson star -- neutron star collisions with orbital angular momentum}

\begin{figure}[t]
\centering
\includegraphics[width=0.95\textwidth]{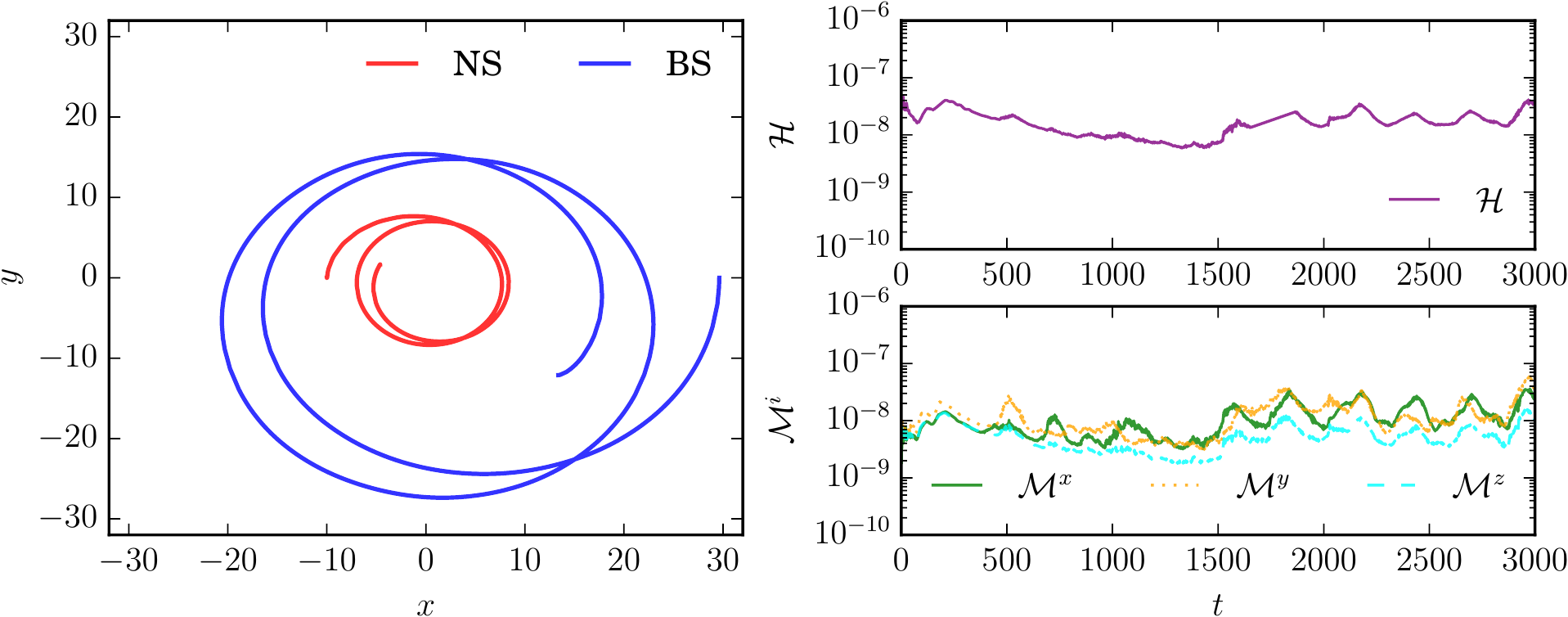}
\caption{Left panel: inspiral tracks of the BS (blue) and the NS (red). 
We compensate for a constant offset and linear drift and move the center of 
mass approximately to the origin. Right panels: 
The Hamiltonian constraint (top panel) and the 
$x$,$y$, and $z$-component of the 
momentum constraint.}
\label{fig:BSNS}
\end{figure}

As a final showcase of the code's abilities we present the merger of a BS and NS
where both stars have initial angular momentum, i.e., are orbiting.
For the particular example we show the tracks of the BS (blue) and NS (red) 
in Fig.~\ref{fig:BSNS}. 
We have chosen an initial amplitude of the scalar field 
of $0.08$ which results in a mass ratio of $\approx 2$ considering 
our standard $1.35$ SLy NS.  
At the current stage, we simply boost the two stars, solve the constraint equations, 
and adjust the NS's density profile according to Eq.~\eqref{eq:rho_update}.
Note that we correct in the plot for center of mass drifts caused by 
a non-zero initial ADM-linear momentum. 
Further tuning of the initial conditions will allow for longer simulations without center 
of mass drifts and eccentricity - the simulation presented here is just meant 
as a proof-of-principle for the newly implemented routines. 
The right hand side of Fig.~\ref{fig:BSNS} shows the Hamiltonian constraint in the top panel and the 
components of the momentum constraint in the bottom panel. 
We find no noticeable difficulty and the constraints remain small throughout the simulation.
A detailed study of the emitted GW signal and comparisons
with BS-BS and NS-NS setups is left for future work.

\section{Summary}
\label{sec:conclusion}

The recent detections of GWs by the advanced LIGO and Virgo detectors
are consistent with the expected GWs emitted by the collisions
of binary black holes and binary neutron stars.
However, it has not yet been ruled out that exotic objects such as
Boson, Axion, or Proca stars were involved.
To prepare for such a study and understand the strong field regime of the
(exotic) compact binary coalescence, full 3D numerical
relativity simulations are required.
To increase the range of systems within reach of full 3D numerical relativity
simulations we extended the \BAM code with the necessary routines for
the simulation of scalar fields coupled to the metric evolution.
This enables the study of compact objects characterized by (complex) scalar fields
with the merger of black holes and neutron stars.
To test the newly implemented routines we performed
intensive studies of single star spacetimes.
We find generally that \BAM is able to stably evolve
BSs with different potential types
(free-field and solitonic). However, the time dependence
of the scalar field variables seems to require
a smaller Courant-Friedrichs-Lewy factor than common
neutron star or black hole simulations employ.
Despite this adjustment, the default gauge choices
and numerical methods employed for the simulation
of black holes and neutron stars also allow 
for accurate boson star simulations.
We tested our numerical method by simulating
the head-on collision of two boson stars, and 
obtained clean second order convergence of the GW signal
even after the merger of the individual stars.
We also find good agreement between the \BAM results
and independent evolutions using the SpEC and GRChombo codes.

In preparation for future work, 
we presented the first 3D numerical relativity simulations of 
mixed binaries consisting of a boson star 
and a neutron star system. These systems are of great 
interest since they (i) could explain the GW signal of compact binaries; 
(ii) would allow for electromagnetic counterparts due to 
the presence of baryonic matter; (iii) if detected they would 
place constraints on the properties of the scalar fields, giving evidence for
new light, bosonic fundamental fields and potentially shedding 
light on dark matter candidates. 

Studying three different head-on configurations, 
we find that depending on the mass of the boson star 
the merger remnant can either be a black hole or a neutron star 
surrounded by a bosonic cloud. For these different cases the 
emitted amplitude of the GW signal can vary by orders of magnitude.

Finally, we showcase \BAM's ability to model mixed inspiraling systems 
by presenting a boson star -- neutron star merger configuration
with orbital angular momentum and leave detailed simulations for future work. 

\ack
  We are particularly thankful to J.~Niemeyer for stimulating this 
  work and for helpful and productive comments through the 
  project. 

  We thank all developers of the BAM code
  and the CoRe Collaboration for enabling us to use BAM 
  for the simulation of compact objects. In particular 
  we are grateful for guidance and support 
  from S.~Bernuzzi and B.~Br\"ugmann.
  We also thank the developers of the SpEC code, as well as 
  the entire SXS collaboration, for their help with the SpEC simulations
  and we thank the developers of the GRChombo code for their help
  and support with the GRChombo simulations. 
  
  TD is supported by the European Union’s Horizon
  2020 research and innovation program under grant
  agreement No 749145, BNSmergers.
  TD also acknowledges the hospitality and discussion with the 
  Cosmology group of the Astrophysics Department of the University of 
  G\"ottingen. 
  Computations have been performed
  on the supercomputer SuperMUC at the LRZ
  (Munich) under the project number pr48pu,
  the compute cluster Minerva of the Max-Planck
  Institute for Gravitational Physics,
  and the GWDG cluster in Goettingen.

\appendix

\section{Conversion to Planck units}
\label{sec:appx}

Finally, we also wish to put our results in the context of Planck units.
When converting to Planck units, it is preferable to think of the code units in terms a length scale $L$,
which corresponds to the Schwarzschild radius of a solar mass BH, rather than a mass scale $M$ .
Then all lengths are expressed in terms of $L$ and times correspond to $L / c$.
This use of length as the fundamental quantity is suggested in the Appendix to Wald \cite{Wald:1984rg},
but in practice the use of mass scales is far more common in NR applications.

However, one needs to take care when converting masses, most notably in considering particle masses,
in which the $\hbar$ must be reintroduced.
Having set $G=c=1$ and $M_\odot=1$, this fixes our mass scale according to the solar mass,
which implies $\hbar \neq 1$.
The boson ``mass" $\mu$ which appears in the potential function $V(\phi) = \mu^2 |\phi|^2$ is the quantity
\begin{equation}
\mu = \frac{mc}{\hbar} ~,
\end{equation}
where $m$ is the usual particle mass in kg or equivalently $\rm eV/c^2$.
$\mu$ thus has dimensions $[L^{-1}]$ and to convert it to SI units one must multiply by $\frac{c^2}{G M_\odot}$.
From this the boson mass is obtained as $m = \hbar \mu / c$, such that a value of $\mu=1$ in code units corresponds
to a particle mass of $m =1.3 \times 10^{-10} ~ {\rm eV/c^2}$, which is in the range of possible QCD axion masses.

\section*{References}

\bibliographystyle{iopart-num}
\bibliography{paper20180718.bbl}

\end{document}